\documentclass[onecolumn,usenatbib,usegraphicx]{mn2e}

\title[Stochastic Electron Acceleration in TeV SNRs]
{Stochastic Electron Acceleration in the TeV Supernova Remnant RX
J1713.7-3946: The High-Energy Cut-off}
\author[Z. H. Fan et al.]{Zhonghui Fan$^{1}$\thanks{E-mail: fanzh@ynu.edu.cn},
Siming Liu$^{2}$\thanks{E-mail: sliu@astro.gla.ac.uk} and
Christopher L. Fryer$^{3,4}$\thanks{E-mail: fryer@lanl.gov}\\
$^1$Department of Physics, Yunnan University, Kunming
650091, Yunnan, China\\
$^2$Department of Physics and Astronomy, University of Glasgow,
Glasgow, G12 8QQ, UK\\
$^3$Los Alamos National Laboratories, Los Alamos, NM 87545\\
$^4$Physics Department, University of Arizona, Tucson AZ 85721}

\begin{document}

\date{}

\pagerange{\pageref{firstpage}--\pageref{lastpage}} \pubyear{0000}

\maketitle

\label{firstpage}

\begin{abstract}
In the leptonic scenario for TeV emission from a few well-observed
shell-type TeV supernova remnants (STTSNRs), very weak magnetic
fields are inferred. If fast-mode waves are produced efficiently in
the shock downstream, we show that they are viable agents for
acceleration of relativistic electrons inferred from the observed
spectra even in the subsonic phase, in spite that these waves are
subject to strong damping by thermal background ions at small
dissipation scales. Strong collisionless non-relativistic
astrophysical shocks are studied with the assumption of a constant
Aflv\'{e}n speed in the downstream. The turbulence evolution is
modeled with both the Kolmogorov and Kraichnan phenomenology.
Processes determining the high-energy cutoff of nonthermal electron
distributions are examined. The Kraichnan models lead to a shallower
high-energy cutoff of the electron distribution and require a lower
downstream density than the Kolmogorov models to fit a given
emission spectrum. With reasonable parameters, the model explains
observations of STTSNRs, including recent data obtained with the
{\it Fermi} $\gamma$-ray telescope. More detailed studies of the
turbulence generation and dissipation processes, supernova
explosions and progenitors are warranted for better understanding
the nature of supernova shocks.
\end{abstract}


\section{Introduction}

The acceleration of cosmic rays up to $\sim 10^{15}$ eV has been
attributed to supernova explosions and TeV emission is expected from
the remnants \citep{gp76, lc83, r08, b09}. The standard diffusive
shock particle acceleration (DSA) model has been successful in
explaining emissions from most supernova remnants \citep{e79c, b87,
k99, z07, v09, r08}. Investigations of acceleration by a spectrum of
turbulent plasma waves, the so-called stochastic particle
acceleration (SA), also have a long and resilient history
\citep{s75, l77, a79, e79, bt83, cs84, p88, bt93, a00, pl04, c06,
l08}. Most authors prefer the use of relativistic leptons to account
for the nonthermal radio, X-rays, and TeV emissions from the
remnants. The TeV emission has also been attributed to energetic
protons and ions \citep{a06, m09, f09, z10}.

Although the DSA can naturally give a universal power-law energetic
particle distribution with the spectral index determined by the
shock compression ratio for a linear model, it requires well-defined
shock structure and efficient scatter of high-energy particles by
small-scale turbulence \citep{b78}. The initial acceleration of
low-energy particles to a high enough energy for the shock to be
effective, the so-called injection problem, is likely due to the SA
by turbulence.
\citet{l77} showed that the SA by small-scale Alfv\'{e}n waves can
be more efficient than the first order Fermi acceleration by shocks.
\citet{a79} derived approximate diffusion coefficients and showed
that stochastic interactions of particles with a spectrum of plasma
waves can lead to efficient particle acceleration. Over the past few
decades, the SA has also been explored for broader astrophysical
applications \citep[e.g.,][]{e79a, e79b, cs84, bmn92, mlm96, sm98,
pl04, yl04, f07, pb08}.
The essential challenges to the SA model are a self-consistent
treatment of the nonlinear turbulence spectral evolution and the
requirement of the same energy dependence of the acceleration and
escape timescales for the production of a power-law particle
distribution \citep{bmn92, bld06, p88}. Recently Gibbsian theory has been
generalized to account for power-law distributions in marginally
stable Gibbsian equilibria \citep{tj08}. It remains to be shown how
the physical processes of the SA are related to the ordering
parameter $\kappa$ of this statistics.  Most previous studies assume
isotropic magnetohydrodynamic (MHD) waves for the turbulence
\citep{l77, a79, e79b, mlm96}. The anisotropy of the turbulence
caused by cascade and damping processes has been considered recently
\citep{g95, cha03, yl04, c06, j09}. In particular, \citet{l08} show that the
scatter and acceleration rates of charged relativistic particles by
fast-mode waves in a high-$\beta$ plasma may be much higher than
those given by the standard quasi-linear theory with an isotropic
wave power spectrum.

Over the past few years, detailed radio, X-ray, $\gamma$-ray, and
TeV observations of a few shell-type TeV supernova remnants
(STTSNRs) pose several challenges to the classical DSA model in the
hadronic scenario, where the TeV emission is produced through
neutral pion decays induced by proton-proton scatter \citep{a06,
a07, t08, funk09}. Besides requiring efficient amplification of the
magnetic field in the upstream plasma and a good correlation between
the magnetic field and background plasma density \citep{p08, f09,
m09}, the model also implies a cosmic ray energy of $\sim 10^{51}$
ergs for each remnant and an electron acceleration efficiency more
than 4 orders of magnitude lower than the ion acceleration
efficiency \citep{b08}. The high density of the upstream plasma in
the model also implies significant thermal X-ray emission from the
downstream, which may exceed the observed upper limit \citep{c04}. A
very hard proton spectrum is also required to fit the $\gamma$-ray
spectrum obtained recently with the {\it Fermi} $\gamma$-ray
telescope \citep{funk09}. Although these kinds of remnants may be
atypical, detailed modeling can still have profound implications for
our understanding of supernova shocks \citep{b09, z10}.

In a previous paper, we showed that the SA of electrons by turbulent
plasma waves in the shock downstream might naturally explain these
observations \citep{l08}. Turbulence is expected given that the size
of these remnants are many orders of magnitude larger than the
dissipation scale of the ion inertial length \citep{d91, j09}. The DSA
model proposes that particle acceleration occurs directly and
predominantly at the ion inertial length or gyro-radius \citep{r08}.
This requires the absence of instabilities over a large range of
spatial scales, which is highly idealized. \citet{gj07} showed that
density fluctuations in the upstream can be amplified significantly
by shock waves, resulting in strong turbulence in the downstream.
Magnetized turbulence appears to be a more generic and natural
energy dissipation channel than the short-length-scale shock fronts
(SF).

With the leptonic model, the TeV emission is mostly produced by the
inverse Comptonization of the cosmic microwave background radiation
by TeV electrons \citep{p06}. The magnetic field required to
reproduce the observed X-ray flux by the same TeV electrons through
the synchrotron process implies a spectral cutoff in the hard X-ray
band, which is in agreement with observations. The SA model also
requires a much lower gas density than the DSA model, which not only
explains the lack of thermal X-ray emission from the shell of the
remnants, but also reduces the energetics of the supernovae. The
required turbulence generation scale is comparable to the size of
the observed X-ray filaments as well \citep{d91, u07}. The fast
variability of small X-ray features may be attributed to rapid spatial
diffusion of high energy electrons \citep{l08}.
As we will show in
this paper, the agreement between preliminary results from {\it Fermi}
observations and the model prediction is also impressive
\citep{funk09}.

The SA by fast-mode waves has been studied by several authors
\citep{bt83, p88, bt93, mlm96, sm98, yl04, c06, l06}. Both resonant
and nonresonant interactions have been considered. In this paper, we
consider the nonresonant acceleration by compressional waves first
studied by \citet{bt83, p88}. Compared with these original studies,
our model has several distinct features: 1) most of the dissipated
fast-mode turbulence energy is absorbed by thermal background ions;
2) the residual fast-mode waves in the dissipation scales propagate
along local magnetic fields and preferentially accelerate electrons
in the background plasma; 3) the plasma physics processes in the
dissipation range in principle may lead to a self-consistent
treatment of the electron injection process at low energies; 4) the
scatter mean-free-path of relativistic electrons, which is a free
parameter in most of the previous studies, is determined by the
characteristic length of the magnetic field, which in a high-$\beta$
plasma is reduced by strong turbulence motions significantly; 5) the
high-energy cutoff of the particle distribution is tied to the
characteristic length of the magnetic field.

In this paper, we first discuss the SA by decaying turbulence in
general and show that fast-mode waves may account for observations
of a few STTSNRs (Section \ref{TS}). In Section \ref{shock}, we
present the structure of the downstream turbulence with both the
Kolmogorov and Kraichnan phenomenology for the turbulence cascade.
The transit-time damping (TTD) by the thermal background particles
of compressional fast-mode turbulence is considered in the
dissipation range. The stochastic acceleration of relativistic
electrons by fast-mode wave turbulence in the subsonic phase is
discussed in Section \ref{SA}, where physical processes determine
the acceleration of the highest energy electrons are discussed. The
models are applied to the well-observed TeV SNR RX J1713.7-3946 in
Section \ref{application}. Although the inferred plasma density is
much lower than that in the hadronic scenario, some models still
have significantly higher densities than that derived from {\it
XMM-Newton} observations \citep{c04}.
In Section \ref{disc}, we discuss how the density may be further
reduced by considering the turbulence generation processes and
first-order Fermi acceleration in the supersonic phase.
Conclusions are drawn in Section \ref{con}.

\section{General Constraints on the Particle Acceleration}
\label{TS}

According to the classical Fermi mechanism \citep{f49}, the particle
acceleration rate is determined primarily by the scatter
mean-free-path $l$ and the velocity of the scatter agents $u$
\citep{b87, bmn92}. Very general constraints can be obtained on the
nature of these processes when applying this mechanism to specific
observations. For example, \citet{e79a} showed that the acceleration
process must be selective in the sense that only a fraction of the
background particles are accelerated to very high-energies.
Otherwise, these stochastic interactions likely lead to plasma
heating instead of a very broad energy distribution of accelerated
particles as frequently observed in dynamically evolving
collisionless astrophysical plasmas. For particle acceleration in
solar flares, \citet{e79b} argued that this selective acceleration
could be achieved in the energy domain (i.e., the frequency domain
for the waves) through cyclotron resonances of particles with a
spectrum of cascading plasma waves. In the DSA model, the particle
acceleration at low energies and the shock structure determine the
efficiency of different particle species \citep{e79a}. In the
presence of a magnetic field, selective energization may also be
realized in the domain of the particle pitch-angle and/or wave
direction angle with respect to the magnetic field \citep{bl08}.

Further constraints can be put on the SA by a spectrum of
turbulence. Given the small gyro-radii of charged particles in
magnetized astrophysical plasmas, charged particles couple strongly
through the magnetic field. As a result, the turbulence responsible
for the SA will decay as the energy carrying plasma being carried
away from the source region of the turbulence by large scale flows
and/or magnetic fields for a high and/or low $\beta$ plasma,
respectively. This is the case for the SA in a shock downstream with
a high value of the plasma $\beta$, where the turbulence is generated
at the SF and its intensity decreases as the flow moves away from the
SF \citep{l08}. We next discuss constraints on such a particle
acceleration scenario.

In the {\it Kolmogorov} phenomenology for the turbulence cascade
\citep{k41}, the free energy dissipation rate is given by
\begin{equation}
Q \equiv C_1\rho u^3/L\,,
\end{equation}
where $C_1$ is a dimensionless constant, $\rho$  is the mass
density, and $u$ and $L$ are the eddy speed and the turbulence
generation scale, respectively. The eddy turnover speed and time at
smaller scales are given respectively by
\begin{eqnarray}
v^2_{edd}(k)&\equiv&4\pi
W(k)k^3\propto k^{-2/3}\,,\\
\tau_{edd}(k) &\equiv& (kv_{edd})^{-1}= (4\pi W
k^5)^{-1/2}\propto k^{-2/3}\,,
\end{eqnarray}
 where
\begin{equation}
W(k) = (u^2/4\pi)L^{-2/3} k^{-11/3}= (4\pi)^{-1} (
Q/C_1\rho)^{2/3} k^{-11/3}\propto k^{-11/3}\, \label{kol}
\end{equation}
is the isotropic turbulence power spectrum, $k=1/l$ is the wave
number and $l$ is the eddy size. From the three-dimension Kolmogorov
constant $C \simeq 1.62$ \citep{yz97}, we obtain $C_1 =  C^{-3/2} =
0.485\,.$ At the turbulence generation scale $L=1/k_m$, $v_{edd}=u$,
$Q= C_1\rho [(4\pi W)^3k^{11}]^{1/2}= C_1\rho v^2_{edd}(k)/
\tau_{edd}(k)\,,$ and the total turbulence energy is given by $\int
W(k) 4\pi k^2 {\rm d} k = (3/2) u^2\,.$ The turbulence decay time is
therefore given by $\tau_d \equiv {\rm d} t/{\rm d}\ln(u)=
3\tau_{edd}(k_m)/C_1\,,$ where $t$ indicates the time, i.e., eddies
decay after making $3/(2\pi C_1)\sim 1$ turn.

We are interested in the acceleration of particles through scattering
randomly with heavy scatter centers with the corresponding
acceleration time given by \citep{b87, bmn92}
\begin{equation}
\tau_{ac} = \tau_{sc} [3 v^2/v^2_{edd}(k)]\,, \label{taukol}
\end{equation}
where
\begin{equation}
\tau_{sc}= (kv)^{-1}=l/v \label{tsckol}
\end{equation}
is the scatter time, $v$ is the particle speed, and we have assumed
that the scatter mean free-path is equal to $l$. For the above
isotropic Kolmogorov turbulence spectrum, $\tau_{ac}(k) = 3
v/(4\pi Wk^4)\propto k^{-1/3}\,.$ To have significant stochastic particle
acceleration, the acceleration time $\tau_{ac}(k)$ should be shorter
than the turbulence decay time $\tau_d$, which implies $u^2>C_1 v
v_{edd}(k)\,.$ So, in general, the SA is more efficient at smaller
scales. The onset scale of the SA is given by $k_c = (C_1v/u)^3
k_m\,$, where $\tau_{ac}=\tau_d$. Therefore, to produce energetic
particles with a speed of $v$ by a Kolmogorov spectrum of scatter
centers, the turbulence must have a dynamical range greater than
\begin{equation}
D_{Kol}=(C_1v/u)^3\,.
\end{equation}
This acceleration process corresponds to the acceleration by
incompressional motions studied by \citet{bt83}.

In the {\it Iroshnikov---Kraichnan} phenomenology \citep{I63, k65, j09},
the turbulence cascade rate is suppressed by the wave propagation effect
by a factor of $v_F/v_{edd}$, where $v_F$ (independ of $k$) is
the wave group speed:
\begin{equation}
C_1\rho v_{edd}^3/(v_F\tau_{edd}) = Q\,.
\end{equation}
Then we have
\begin{eqnarray}
Q&=&C_1\rho u^4/(Lv_F)\,, \\
W(k) &=& (u^2/4\pi)k_m^{1/2}k^{-7/2}\,,\\
v_{edd}&= & u (k/k_m)^{-1/4}\,, \\
\tau_{edd}& = & u^{-1}k_m^{-1/4}k^{-3/4}\,, \\
\tau_{ac}&=& (3 v/u^2)(k_m k)^{-1/2}\,, \label{taukra}
\end{eqnarray}
and the turbulence decay time is given by
\begin{equation}
\tau_d=3\tau_{edd}(k_m)v_F/(C_1u)\,,
\end{equation}
where the wave speed $v_F\gg u$. To have significant acceleration
through scatter with the eddies, the dynamical range of the
turbulence must be greater than
\begin{equation}
D_{IK}=(C_1v/v_F)^2\,,
\end{equation}
which is much less than $D_{Kol}=(C_1v/u)^3$.

The resonant interactions of particles with waves may be more
effective in accelerating particles in this case than the
interactions with eddies. For a wave phase speed $v_F$ independent
of $k$ and a wave spectrum of $W(k)=v_F^2(4\pi
k_m^3)^{-1}(k/k_m)^{-\delta}$, where $\delta$ is the wave spectral
index, the standard quasi-linear theory gives a scatter time of
$\tau_{sc}\simeq (L/v) (r_g k_m)^{4-\delta}=
[v_F^2/v_{edd}(r_g^{-1})^2] r_g/v$ for particles with a gyro-radius
of $r_g=k_0^{-1}$. And the acceleration time is given by $\tau_{ac}
\simeq  3 \tau_{sc} v^2/v_F^2\simeq (3v^2/v_{edd}(k_0)^{2})/(vk_0)$,
which is essentially the same as equations (\ref{taukol}) and
(\ref{tsckol}). Although the acceleration rate depends on the phase
speed of these waves, the scatter rate is proportional to the
intensity of waves in resonance with the particles. So the
acceleration is not enhanced by these kinds of resonant
interactions. In the presence of magnetic fields, the stochastic
acceleration of charged particles can be very efficient
\citep{cha03}. \citet{c06} reviewed several mechanisms by MHD waves,
which can accelerate particles within the turnover time of large
scale eddies for appropriate particle spatial diffusion
coefficients. Some of these mechanisms likely result in plasma
heating instead of particle acceleration as the particle
energization processes are not selective. The corresponding waves
are also likely subject to efficient damping by the thermal
background particles. There are other mechanisms, which
preferentially accelerate energetic particles.

For isotropic acoustic wave turbulence with a wave phase speed $v_F$,
\citet{p88} showed that the stochastic acceleration timescale of energetic
particles with a spatial diffusion coefficient
$D=\tau_{sc}v^2/3 > v_F/k_d$ is given by (see also \citet{bt83})
\begin{equation}
\tau_{ac} \simeq [3 +\xi(\delta)
 (Dk_m/v_F)^{3-\delta}]D/u^2\,, \label{tac0}
\end{equation}
where $3u^2/2$ is the overall wave intensity, $k_m$ indicates the
wave generation scale, and below the turbulence dissipation scale
$l_d=1/k_d$ the wave intensity is negligible. $\xi(\delta)$ depends
on $\delta$ and is on the order of unity (e.g., $\xi(11/3)\simeq
3.0$). The results for fast and slow diffusion have been combined
here approximately to give a unified expression \citep{p88, c06}.
For $\delta<4$, $\tau_{ac}$ increases monotonically with $D$. To
have significant acceleration, the minimum acceleration timescale
$\tau_{ac\rm min}= \tau_{sc}v^2/u^2$ must be less than $\tau_d =
3Lv_F/(C_1u^2)$. Then the scatter mean-free-path of the particles
$\tau_{sc}v$ must be shorter than
\begin{equation}
L/D_{AI} = 3Lv_F/(C_1v) = 3L/D_{IK}^{1/2}\,.
\end{equation}
It is also possible that the dynamics of the turbulence cascade is
not affected by the wave propagation, which only enhances the
particle acceleration rate. Then $\tau_{d} = 3L/(C_1u)$, the dynamical
range required for
the SA to be significant is given by
\begin{equation}
D_{AK} \equiv C_1v/(3u) = D_{Kol}^{1/3}/3\,.
\end{equation}

Several STTSNRs have been observed extensively in the radio, X-ray,
and TeV bands. X-ray observations with the {\it Chandra}, {\it
XMM-Newton}, and {\it Suzaku}, and TeV observations with the HESS
have made several surprising discoveries that challenge the DSA
model in the hadronic scenario \citep{a06, a07, l08,t08}. The
leptonic scenario, on the other hand, is relative simple except that
the electron acceleration mechanism needs to be addressed
\citep{z07,l08,v09}. The TEV SNR RX J1713.7-3946 is about $T_{\rm
life} =1600$ years old \citep{w97} with a radius of $R\simeq 10$ pc
and a distance of $D \simeq 1$ kpc.  By fitting its broadband
spectrum with an electron distribution of $f\propto \gamma^{-p} \exp
-(\gamma/\gamma_c)^{1/2}$ (the dashed lines in Fig. \ref{spec}), we
find that $p = 1.85$, $B = 12.0\; \mu$G, $ \gamma_c m_e c^2 = 3.68$
TeV, and the total energy of relativistic electrons with the Lorentz
factor $\gamma>1800$ $E_e=3.92 \times 10^{47}$ erg.

The X-ray emitting electrons have a gyro-radius of $r_g\simeq
10^{15}$ cm, which should be shorter than the particle scatter
mean-free-path. To produce these electrons through the SA, the
turbulence must be generated on scales greater than $D_{Kol}r_g$,
$D_{IK}r_g$, $D_{AK} r_g$, and $D_{AI} r_g$ for the non-resonant
Kolmogorov, Kraichnan, acoustic Kolmogorov, and Kraichnan
phenomenology, respectively. For STTSNRs, $u\sim v_F\sim 0.01c$,
$D_{Kol}r_g\sim 10$ kpc, which is much greater than the radii of the
remnants. The SA by eddies with a Kolmogorov spectrum is therefore
insignificant. The standard quasi-linear theory also predicts
negligible SA. $D_{IK}r_g\sim 30$ pc, which is also too thick.
$D_{AI} r_g\sim D_{AK} r_g\sim 0.03$ pc, which is much greater than
the ion inertial length and may result from Kelvin-Helmholtz
instabilities or cosmic ray drifting upstream \citep{b78,
m99,gj07,n08}. Therefore if relativistic electrons from the STTSNRs
are accelerated through the SA, they may be energized by high-speed
compressional plasma waves.
Previously \citet{e79a} showed that the non-selective acceleration
as given by equations (\ref{taukol}) and (\ref{taukra}) leads to
plasma heating instead of particle acceleration. Here we give a
different argument against these acceleration processes. The
acceleration timescale in the following is given by equation
(\ref{tac0}) with $D$ depending on the wave spectrum and particle
energy.

The DSA models are usually favored over the SA models for two main
reasons: 1) the DSA can naturally produce a power-law particle
distribution; 2) the DSA corresponds to a first-order Fermi
mechanism and is presumably more efficient than the SA, which
corresponds to a second-order Fermi mechanism. While how shocks
produce power-law high-energy particle distributions is relatively
well understood, the second reason appears to be a misconception. It
is true that the ratios of the acceleration and scatter rates are
proportional to the first and second power of the speed ratio of the
scatter agent and the particle for the DSA and SA, respectively. But
in the DSA models, there are two scatter processes: scatter of
particles by turbulence, which causes the particle diffusion, and
the particle crossing of the SF due to this diffusion. The SF
crossing rate is about a factor of $v/u$ lower than the particle
scatter rate by the turbulence for a shock speed of $\sim u$
\citep{lc83}. As a result, the acceleration rates of the DSA and SA
models are both proportional to $u^2/lv$. Therefore the SA is not
necessarily less efficient than the DSA. The SA at small scales can
also be enhanced by high speed kinetic plasma waves \citep{p97,
pl04}. Of course, the speed of the scatter agent accessible to a
particle may depend on the scatter mean-free-path $l$ in the SA. The
stochastic acceleration rate can be very low if $u^2$ decreases
quickly with the decrease of $l$. For the DSA and SA by large-scale
acoustic waves in the shock downstream, the speed of the scatter
agent is proportional to the shock speed. One just needs to reduce
the scatter mean-free-path $l$ to enhance the acceleration rate. As
we will show in Sections \ref{shock} and \ref{SA}, considering the
competition of isotropic cascade with anisotropic damping of
fast-mode waves through the transit-time-damping process, the waves
in the dissipation range propagate along local magnetic field lines
with a spectrum index of $2$. In a high-$\beta$ plasma, the damping
is mostly caused by ions. The residual parallel propagating waves
however preferentially accelerate electrons. Enegetic particles are
also subject to efficient acceleration by large-scale fast-mode
(acoustic) waves with energy independent acceleration and scatter
rates, giving rise to power-law particle distributions in the
steady-state. At even higher energies, the gyro-radius of the
particles exceeds the characteristic length of the magnetic field,
which also corresponds to the dissipation scale, the spatial
diffusion coefficient $D$ increases quickly with energy. However,
due to interactions with large-scale turbulence, both the
acceleration and spatial diffusion timescales vary gradually with
$D$, implying a gradual high-energy cutoff.

\section{Shock Structure and Damping of Fast-Mode Waves in the Downstream}
\label{shock}

We consider the relatively simple case, where both the thermal
pressure and magnetic field are negligible in the upstream and the
shock normal is parallel to the plasma flow. The mass, momentum, and
energy fluxes are given, respectively, by $\rho V_0$, $P+\rho
V_0^2$, and $V_0(E + P +\rho V_0^2/2)$, where $V_0$, $P$, and $E$
are the speed, pressure, and energy density of the plasma flow,
respectively. Based on the parameters inferred from the leptonic
scenario for the TeV emission from STTSNRs shown in Section
\ref{TS}, the plasma $\beta$ in the shock downstream is likely high
and the fast-mode wave damping by thermal background is then
dominated by protons and ions \citep{l08}. The Alfv\'{e}n speed
given by $v_A=(B^2/4\pi\rho)^{1/2}$ therefore must be much smaller
than the turbulence speed $u$ near the shock front, where $B$ is the
magnetic field intensity.  For strong non-relativistic shocks with
the shock frame upstream speed $U$ much higher than the speed of the
parallel propagating fast mode waves in the upstream $v_F =
(v_A^2+5v_S^2/3)^{1/2}$, where $v_S^2=P_g/\rho$ is the isothermal
sound speed and $P_g$ is the thermal pressure of the gas, $V_0=U$
and $\rho U^2 \gg P\sim E$ in the upstream. In the downstream, the
pressure and energy density have contributions from the thermal gas
and turbulence: $P = \rho (v_S^2 + u^2)$ and $E = \rho
(3v_S^2/2+3u^2/2)$, where we have assumed that the turbulence
behaves as an ideal gas and ignored possible dynamical effects of
the wave propagation. Then we have \citep{t71}
\begin{eqnarray}
\rho_u U & = & \rho_d U/4\,, \\
\rho_u U^2 & = & \rho_d (U^2/16 + v_S^2 + u^2)\,, \\
\rho_u U^3/2 & = & \rho_d U(U^2/16+5v_S^2+5u^2)/8\,,
\end{eqnarray}
where the subscripts $u$ and $d$ denote the upstream and downstream,
respectively, and we have ignored the effects of the electromagnetic
fields and the thermal energy and pressure in the upstream. Then we
have
\begin{equation}
U^2= 5 v_S^2 + 5 u^2+U^2/16\,. \label{energy}
\end{equation}
This is slightly different from that given by \citet{l08}, where we
assumed that the pressure and enthalpy of the turbulent magnetic
field are given by $\rho_d v_A^2/3$ and $ 5\rho_d v_A^2/6$,
respectively.

The shock structure can be complicated due to the present of
turbulence. We assume that the turbulence is generated isotropically
and has a generation scale of $L$, which does not change in the
downstream. The speeds $v_S$, $v_A$, and $u$ therefore should be
considered as averaged quantities on the scale $L$. $v_A$ depends on
the upstream conditions and/or the dynamo process of magnetic field
amplification \citep{lb00, c00,n08}. Here we assume it a constant in
the downstream. One can then quantify the evolution of other speeds
in the downstream.

For the Kolmogorov phenomenology \citep{z90},
\begin{equation}
{3{\rm d} \rho u^2 \over 2{\rm d} t}= -Q\ \ \ {\rm i.e.,}\ \ \
{3U{\rm d} u(x)^2 \over 8{\rm d} x}  = -{C_1 u(x)^3\over L}\,.
\label{ux}
\end{equation}
Near the SF, we denote the isothermal sound speed and Aflv\'{e}n
speed by $v_{S0}$ and $v_{A0}$, respectively. Then the eddy speed at
the SF is given by $a^{1/2}U/4$ with $a = 3 - 16 v_{S0}^2/U^2$.
Integrate equation (\ref{ux}) from the SF $(x=0)$ to downstream
($x>0$), we then have
\begin{eqnarray}
{u(x)\over U} &=& {1\over 4C_1x/3L+ 4/a^{1/2}}\,, \\
{v_S(x)\over U} &=& \left[{3\over 16}-{1\over
16\left(C_1x/3L+a^{-1/2}\right)^2} \right]^{1/2}\,, \\
{v_F(x)\over U}&=& \left[{5\over 16}-{5\over
48\left(C_1x/3L+a^{-1/2}\right)^2} + {v_A^2\over
U^2}\right]^{1/2}\,.
\end{eqnarray}
\begin{figure}
\begin{center}
\includegraphics[width=0.48\textwidth]{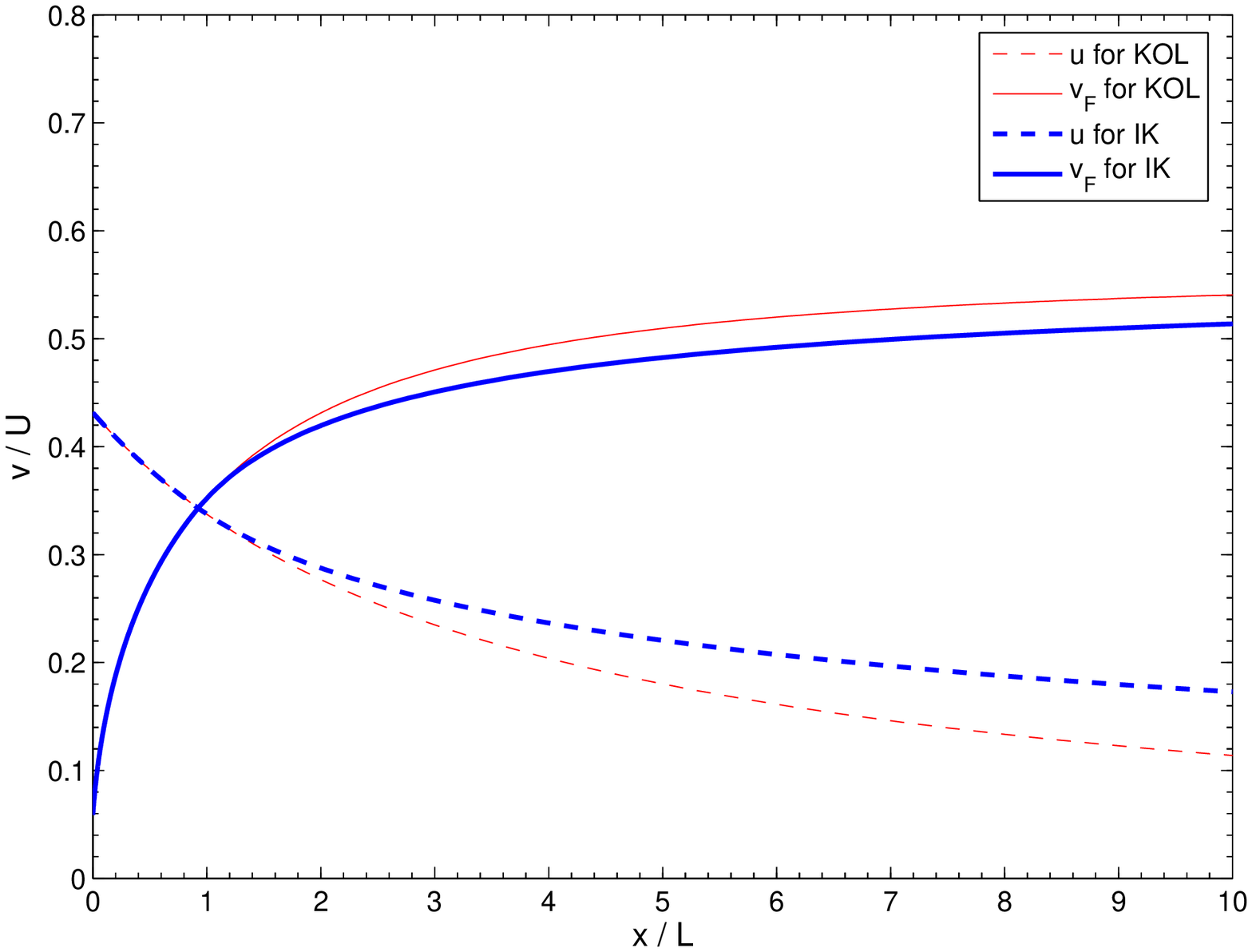}
\includegraphics[width=0.48\textwidth]{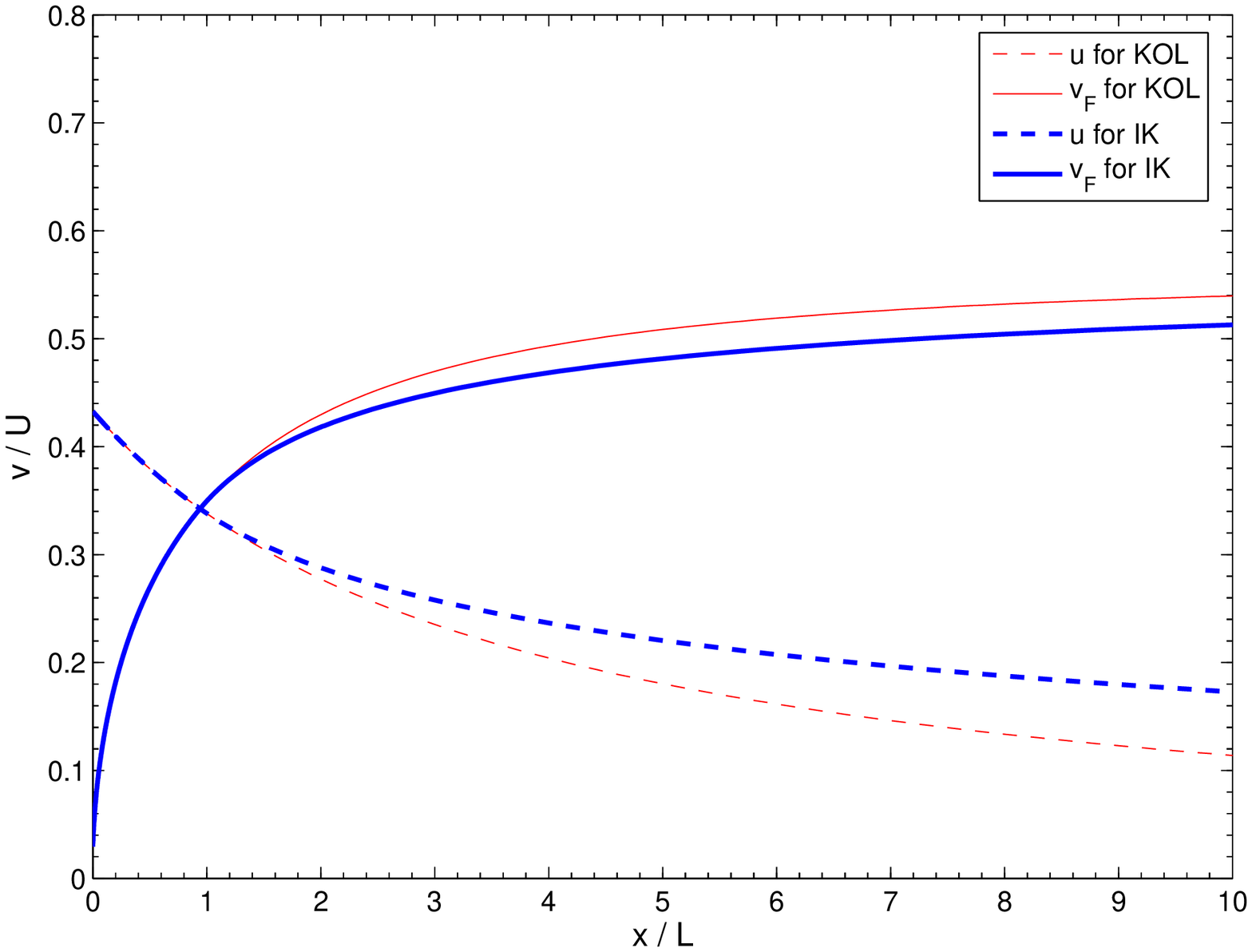}
\end{center}
\caption{Evolution of the eddy speed $u$ and the speed of parallel
propagating fast mode waves $v_F$ in the downstream. The thick and
thin lines are for the Kraichnan with $v_A = 0.036U$ (left) and
Kolmogorov with $v_A = 0.018U$ (right) phenomenology, respectively.
\label{vs} }
\end{figure}

As mentioned in Section \ref{TS}, to produce the observed X-ray
emitting electrons in the STTSNRs through the SA processes,
fast-mode waves needs to be excited efficiently. The MHD wave period
is given by $\tau_F(k)=2\pi /v_Fk$. Then the transition from the
Kolmogorov to Kraichnan phenomenology occurs at the scale, where
$\tau_F(k_t) = 2\pi \tau_{\rm edd}(k_t)$ or $v_F = v_{\rm edd}(k_t)$
\citep{j09}. We then have
\begin{equation}
k_t = (u/v_F)^3k_m\,.
\end{equation}
For $k_t>k>k_m$, the turbulence spectrum is Kolmogorov like given by
equation (\ref{kol}). For $k>k_t>k_m$, the turbulence spectrum in
the inertial range is given by
\begin{equation}
W(k) 
={1\over 4\pi}\left\{ 
\begin{array}{ll}
v_F^{1/2}u^{3/2}k_m^{1/2}k^{-7/2}\ \ \ \ &{\rm for\ IK}\,, \\
u^2 k_m^{2/3} k^{-11/3}\ \ \ \ \ &{\rm for\ Kol}\,.\\
\end{array}
\right.
\end{equation}
Although the turbulence energy exceeds $(3/2)u^2$ when the wave
propagation effect is considered, we still assume that the enthalpy
of the turbulence is given by $(5/2) u^2$ for $v_F<u$ so that
equation (\ref{energy}) and the above solutions for the speed
profiles remain valid.

In the subsonic phase with $v_F>u$,
\begin{equation}
W(k) 
={u^2\over 4\pi}\left\{ 
\begin{array}{ll}
k_m^{1/2}k^{-7/2}\ \ \ \ &{\rm for\ IK}\,, \\
k_m^{2/3} k^{-11/3}\ \ \ \ \ &{\rm for\ Kol}\,.\\
\end{array}
\right.
\end{equation}
and
\begin{eqnarray}
{3U{\rm d}u(x)^2\over 8{\rm d} x}&=& -{C_1 u(x)^4\over Lv_F}\ \ \
{\rm for\ IK}\,, \label{ik}
\end{eqnarray}
where from equation (\ref{energy}) one has $v_F = \left[{5U^2/16}+
{v_A^2-{5u^2(x)/3}} \right]^{1/2}.$ Equation (\ref{ik}) can be
solved numerically to get the speed profiles in the subsonic phase.
Figure \ref{vs} shows the $v_F$ and $u$ profiles with
$v_A=v_{A0}=v_{S0}\ll U$ in the downstream. The thick and thin lines
are for the Kraichnan and Kolmogorov phenomenology, respectively.

In summary, for $k>\max{(k_m, k_t)}$, the turbulence spectrum in the
inertial range is given by
\begin{equation}
W(k) 
={1\over 4\pi}\left\{ 
\begin{array}{ll}
u^{3/2}\min({v_F^{1/2}, u^{1/2}})k_m^{1/2} k^{-7/2}\ \ \ \ &{\rm for\ IK}\,, \\
u^2 k_m^{2/3} k^{-11/3}\ \ \ \ \ &{\rm for\ Kol}\,,\\
\end{array}
\right. \label{IK}
\end{equation}
and for $k_t>k>k_m$, $W(k) = u^2 (4\pi)^{-1} k_m^{2/3} k^{-11/3}$.

The transit-time damping (TTD) of compressional fast-mode waves
starts at the characteristic length of the magnetic field $l_d =
1/k_d\,, $ where the Alfv\'{e}n speed is comparable to the eddy
speed, i.e.,
\begin{equation}
v_A^2 =  4\pi W(k_d) k_d^3 = \left\{
\begin{array}{ll}
\min({v_F^{1/2}, u^{1/2}})
u^{3/2}k_m^{1/2} k_d^{-1/2}\ \ \ \ \ &{\rm for\ IK}\,, \\
u^2 (k_m/k_d)^{2/3}\ \ \ \ \ &{\rm for\ Kol}\,.\\
\end{array}
\right.
\end{equation}
At even larger scales, the vortex motions produce random magnetic
fields comparable with the mean field reducing the scatter
mean-free-path of charged background particles to $l_d$ or even
shorter scales. This trapping of charged particles within a scale of
$l_d$ prevents the TTD on scales above $l_d$. The incompressional
modes are not subject to the TTD and have different spectra
\citep{g95, c06}. In what follows, we only consider the
compressional (fast magnetosonic) wave modes. Then we have
\begin{equation}
k_d = k_m(u^3/v_A^3) \left\{
\begin{array}{ll}
\min(v_F, u)/v_A\ \ \ \ \ &{\rm for\ IK}\,, \\
1\ \ \ \ \ &{\rm for\ Kol}\,.\\
\end{array}
\right. \label{kd}
\end{equation}
Since $v_F>v_{\rm A}$, one has $k_t<k_d$, and damping is negligible
in the regime, where $k_t>k>k_m$.

For a fully ionised hydrogen plasma with isotropic particle
distributions, which is reasonable in the absence of strong
large-scale magnetic fields, the TTD rate is given by \citep{s62, pyl06}
\begin{eqnarray}
&&\Lambda_T(\theta, k) = {(2\pi k_{\rm B})^{1/2}k\sin^2\theta\over
2(m_e+m_p)\cos\theta} \times \nonumber \\
&&\left[\left(T_em_e\right)^{1/2} e^{-{m_e\omega^2\over 2 k_{\rm B}
T_ek_{||}^2}} + (T_pm_p)^{1/2} e^{-{m_p\omega^2\over 2k_{\rm B}
T_pk_{||}^2}}\right] \label{d1}
\end{eqnarray}
where $k_{\rm B}$, $T_e$, $T_p$, $m_e$, $m_p$, $\theta$, $\omega$,
and $k_{||}=k\cos\theta$ are the Boltzmann constant, electron and
proton temperatures and masses,  angle between the wave vector and
mean magnetic field, wave frequency, and parallel component of the
wave vector, respectively. The first and second terms in the
brackets on the right hand side correspond to damping by electrons
and protons, respectively. For weakly magnetized plasmas with
$v_A<v_S$, proton damping always dominates the TTD for
$\omega^2/k_{||}^2\sim v_S^2\sim k_{\rm B} T_p/m_p$. \footnote{For $T_e=T_p$,
the electron damping term dominates when $v_A\ge 1.9\, v_S$.} If
$v_A$ does not change dramatically in the downstream, the continuous
heating of background particles through the TTD processes makes
$T_p\rightarrow (m_p/m_e) T_e$ since the heating rates are
proportional to $(mT)^{1/2}$, where $m$ and $T$ represent the mass
and temperature of the particles, respectively. We see that parallel
propagating waves (with $\sin \theta=0$) are not subject to the TTD
processes and can accelerate some particles to relativistic energies
through resonant interactions. Obliquely propagating waves are
damped efficiently by the background particles. Although the damping
rates for waves propagating nearly perpendicular to the magnetic
field ($\cos\theta \simeq 0$) are also low, these waves are subject
to damping by magnetic field wandering \citep{pyl06}.

\begin{figure}
\includegraphics[height=.27\textheight,width=0.48\textwidth]{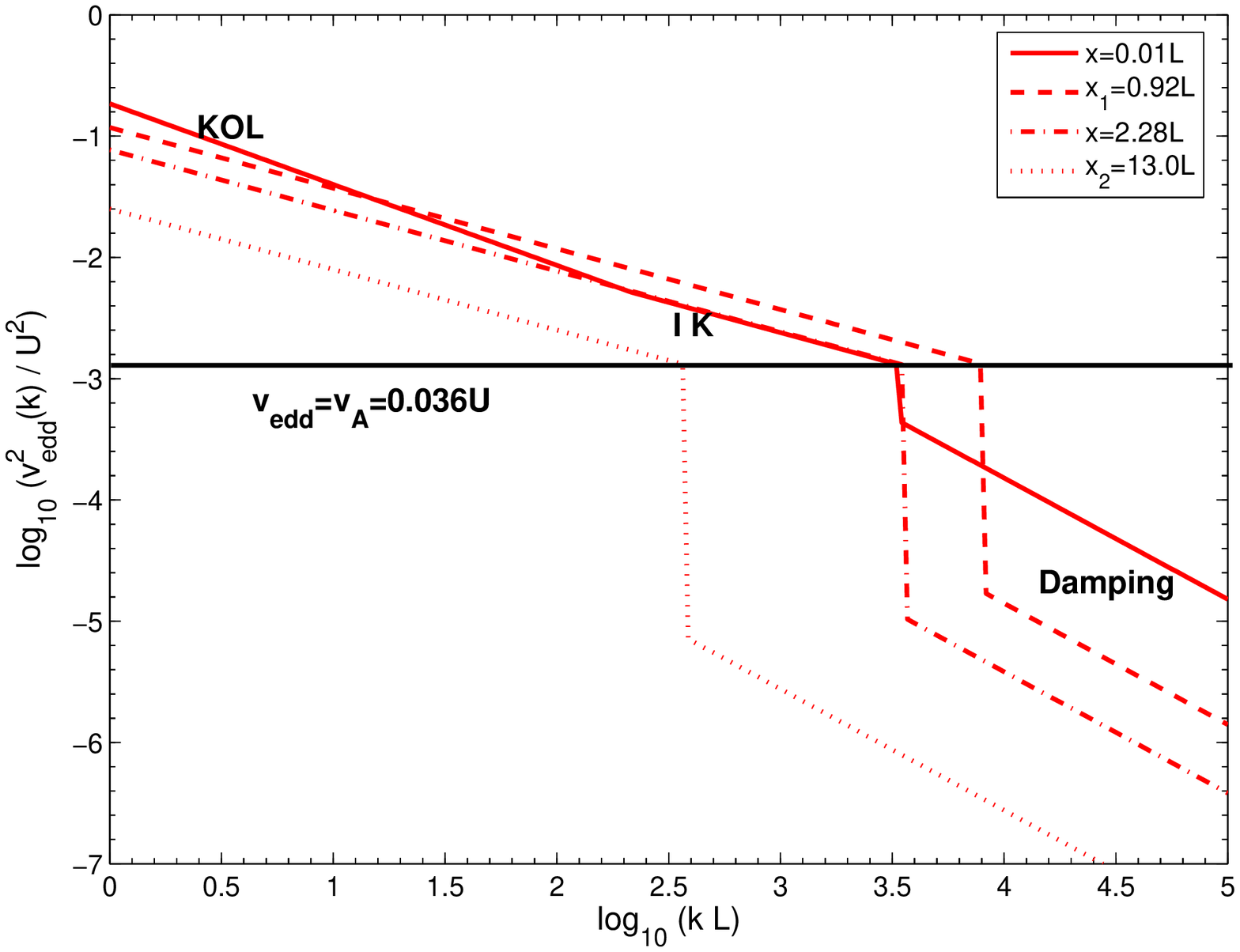}
\includegraphics[height=.27\textheight,width=0.48\textwidth]{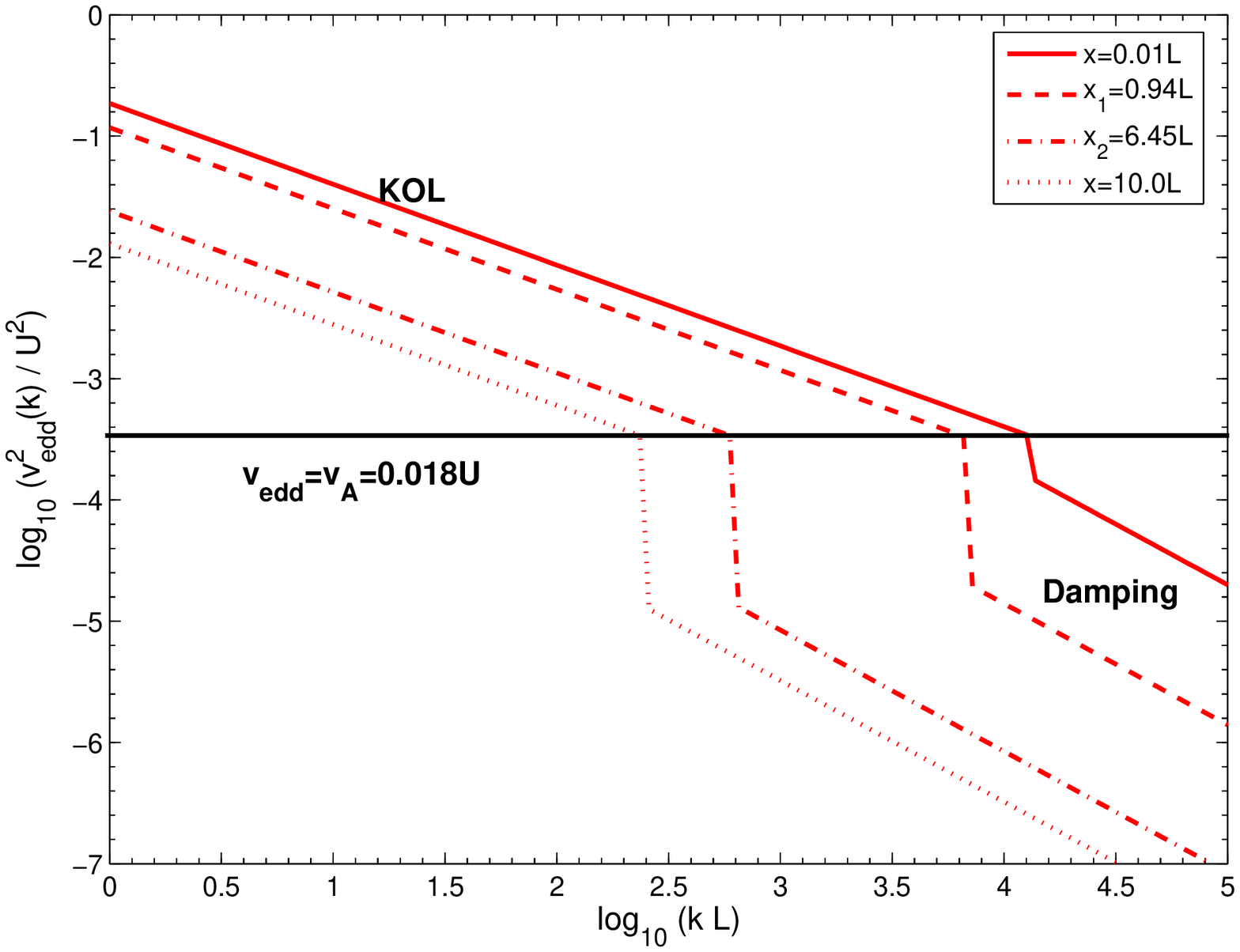}
\caption{ The angular-integrated compressional mode turbulence
spectra $v_{edd}^2(k)$ at several locations in the downstream
indicated in the legend for the speed profiles in Figure \ref{vs}.
In the dissipation range, it corresponds to $k{\cal W}$ instead of
$v_{edd}^2$. The onset of the TTD at $k_d$ causes the discontinuity
for the former. The left panel is for the Kraichnan phenomenology.
The Kolmogorov, Kraichnan, and damping ranges are indicated for the
supersonic phase spectrum with $x=0.010L$. At the other locations,
the turbulence is subsonic and there are only Kraichnan and damping
ranges. The right panel is for the Kolmogorov phenomenology.
\label{vedd}}
\end{figure}

The turbulence power spectrum cuts off sharply when the damping rate
becomes comparable to the turbulence cascade rate \citep{j09}
\begin{equation}
\Gamma =\tau_{edd}^{-1} \left\{
\begin{array}{ll}
\tau_{edd}^{-1}/(\tau^{-1}_F +\tau^{-1}_{edd})\simeq
 \tau_{edd}^{-1}\tau_F \ \ \ \ \ &{\rm for\ IK}\,, \\
1\ \ \ \ \ &{\rm for\ Kol}\,.\\
\end{array}
\right.
\end{equation}
One can define a critical propagation angle $\theta_c(k)$, where
$\Lambda_T(\theta_c, k) = \Gamma(k)$.
Equations (\ref{IK}) and (\ref{d1}) then give
\begin{equation}
{\sin^2\theta_c\over \cos\theta_c} \exp\left(-{v_F^2 \over
2v_S^2\cos^2\theta_c}\right) \simeq \left\{
\begin{array}{ll}
{v_A^2 k_d^{1/2}/(2^{-1/2}\pi^{1/2}  v_S v_F k^{1/2}})\ \ \ \ \ &{\rm
 for\ IK}\,, \\
{v_A k_d^{1/3}/(2^{-1/2}\pi^{1/2}  v_S k^{1/3}})\ \ \ \ \ &{\rm for\ Kol}\,.\\
\end{array}
\right.
\end{equation}
where the electron damping term has been ignored. The
angular-integrated turbulence spectrum in the dissipation range is
therefore given by
\begin{eqnarray}
{\cal W}(k) 
&\simeq& {\theta_c^2(k)\over 2}\left\{
\begin{array}{ll}
u^{3/2}\min({v_F^{1/2}, u^{1/2}})k_m^{1/2} k^{-3/2}\ \ \ \ &{\rm for\ IK}\,, \\
u^2 k_m^{2/3} k^{-5/3}\ \ \ \ \ &{\rm for\ Kol}\,.\\
\end{array}
\right. \nonumber \\
&\simeq& {\exp(5/6)v_A^3 k_d\over 2^{1/2}\pi^{1/2} v_S}k^{-2}\left\{
\begin{array}{ll}
v_A/v_F\ \ \ \ \ &{\rm for\ IK}\,, \\
1\ \ \ \ \ &{\rm for\ Kol}\,.\\
\end{array}
\right. \label{disspec}
\end{eqnarray}
Interestingly the turbulence spectrum is inversely proportional to
$k^2$ in both scenarios. The angular-integrated turbulence
spectra $\int W(k) 2\pi k^2 {\rm d\cos\theta}$ for the velocity
profiles in Figure \ref{vs} at several locations in the downstream
are shown in Figure \ref{vedd}. The discontinuities of the
angular-integrated turbulence spectra are caused by the abrupt onset
of thermal damping at the characteristic length $l_d$ of the
magnetic field. Obliquely propagating fast-mode waves are absorbed
by the thermal background ions at this scale.

\section{Stochastic Electron Acceleration by Fast-Mode Waves in the Downstream}
\label{SA}

In a magnetized plasma, fast-mode waves are likely the agent
responsible for efficient SA of electrons \citep{bt83, bt93, cha03}.
The resonant interactions of particles with fast-mode waves have
been studied by several authors \citep{mlm96, sm98, pl04}. In these
studies, the authors prescribed the wave spectrum with several
parameters and calculated the corresponding Fokker-Planck
coefficients. A self-consistent treatment of the turbulence spectral
evolution in the dissipation range was presented by \citet{yl04} for
a variety of astrophysical plasmas. \citet{l06} considered the
damping of fast-mode waves in a low-$\beta$ plasma and the
application of the SA of relativistic protons in magnetic field
dominated funnels derived from general relativistic MHD simulations
of non-radiative accretion flow around black holes.  Section
\ref{shock} shows that only fast-mode waves propagating parallel to
the local magnetic field can survive the TTD by the thermal ions in
the high-$\beta$ downstream plasma. These waves are right-handed
polarized and resonate preferentially with electrons in the thermal
background and therefore may selectively accelerate electrons to
relativistic energies \citep{pl04}. We next explore the SA of
electrons in the shock downstream in this scenario.

We assume that the turbulence is isotropic from the turbulence
generation scale $L$ down to the dissipation scale $l_d = 1/k_d$,
which depends on the turbulence speed $u$ (Eq. [\ref{kd}]).
Fast-mode waves are excited at the scale $l_F = 1/\max{(k_t, k_m)} =
L/\max{(1, u^3/v_F^3)}$ and are also isotropic down to the
dissipation scale. Due to the TTD, the energy density of fast-mode
waves in the dissipation range is less than the magnetic field
energy density. The resonant scatter rate of energetic particles by
MHD waves is therefore smaller than $v/l_d$. However, for this
high-$\beta$ plasma, particles with a gyro-radius $r_g$ less than
the characteristic length of the magnetic field $l_d$ draft along
magnetic field lines with a scatter mean-free-path $l=v\tau_{sc}\le
l_d$. The scatter mean-free-path of particles with $r_g \simeq l_d$
should be comparable to $r_d\simeq l_d$ since particles with even
higher energies scatter with the magnetic field randomly instead of
performing gyro-motions. This efficient scatter of low-energy
particles by the turbulence magnetic field results from bending of
magnetic field lines by strong turbulence in the inertial range
beyond $l_d$. The resonant wave-particle interactions in the
dissipation range give a much lower scatter rate due to the strong
damping of obliquely propagating fast-mode waves by thermal ions.
Although this scatter through the particle gyro-motion and chaotic
magnetic field struture is not caused by resonances with fast-mode
waves, it determines the spatial diffusion coefficient
$v^2\tau_{sc}/3 \simeq v l_d/3$, which plays essential roles in the
SA by large-scale fast-mode waves \citep{p88}.

The SA in the supersonic phase with $u>v_F$ is not well understood
\citep{a90, bt93}. If we assume the turbulence speed $u$ as the
characteristic speed of the scatter agents and a scatter rate of
$l_d/c$ for relativistic particles with $r_g<l_d$, where $c$ is the
speed of relativistic particles, the acceleration rate will be
comparable to that of the DSA. We also note that the turbulence
speed $u$ is higher than $v_F$ in a narrow region near the SF (Fig.
\ref{vs}) corresponding to the turbulence generation. Such an SA
will be difficult to distinguish from the first-order DSA. Further
downstream, the turbulence speed is lower than $v_F$. We will ignore
the particle acceleration in the supersonic phase and only consider
the acceleration by acoustic waves in the subsonic phase, where
Equation (\ref{tac0}) is approximately applicable \citep{p88}.

We assume $\tau_{sc}=l_d/v=l_d/c$ in the following for relativistic
particles with $r_g\le l_d$. Particles with even higher energies
have a scatter time $\tau_{sc}\ge r_g/c$. The corresponding spatial
diffusion coefficient is given by  $D= \tau_{sc}c^2/3$ \citep{lc83}.
Following \citet{p88}, we have the acceleration timescale of these
particles by a spectrum of fast-mode waves given by
\begin{equation}
\tau_{ac} = \left[{8\pi D\over 9}\int_{k_m}^{k_d}dk {k^4 W(k)\over
         v_F^2+ D^2k^2}\right]^{-1}\,.
\label{tac}
\end{equation}
The fast-mode turbulence also enhances the spatial diffusion of
these particles. The incompressional modes are more efficient in
enhancing the spatial diffusion than compressional modes
\citep{bt93}. To include these effects and partially take into the
effects of energy loss due to adiabatic expansion \citep{bt83,
cs84}, we adopt an effective diffusion coefficient:
\begin{equation}
D_* = D + \chi uL, \label{dstar}
\end{equation}
where $\chi$ is a dimensionless parameter. The escape time of
relativistic electrons from the acceleration region is then given by
\begin{equation}
\tau_{esc} = (k_m^2 D_*)^{-1}=L^2/D_*\,.
\label{tesc}
\end{equation}
Physically, $\chi$ needs to be less than 1, which corresponds to
maximum diffusion caused by turbulence. However, considering the
possible presence of incompressional modes and energy loss due to
adiabatic expansion, the acceleration timescale by compressional
modes will increase since the incompressional modes will carry part
of the turbulence energy. The effect of this increase of
acceleration timescale on the electron distribution can be partially
taken into account by reducing the escape timescale, i.e., by
increasing the effective spatial diffusion coefficient $D_*$. In
what follows, we will treat $\chi$ as a free parameter with $\chi>1$
indicating the presence of incompressional modes and reduction of
acceleration by fast-mode waves. The presence of incompressional
modes will also bring the transonic point closer to the SF. The
acceleration by compressible modes in the subsonic phase can still
be efficient. The details of these processes will depend on the
coupling between compressible and incompressible modes \citep{pb08}
and is beyond the scope of this paper.

\begin{figure}
\includegraphics[height=.27\textheight,width=0.48\textwidth]{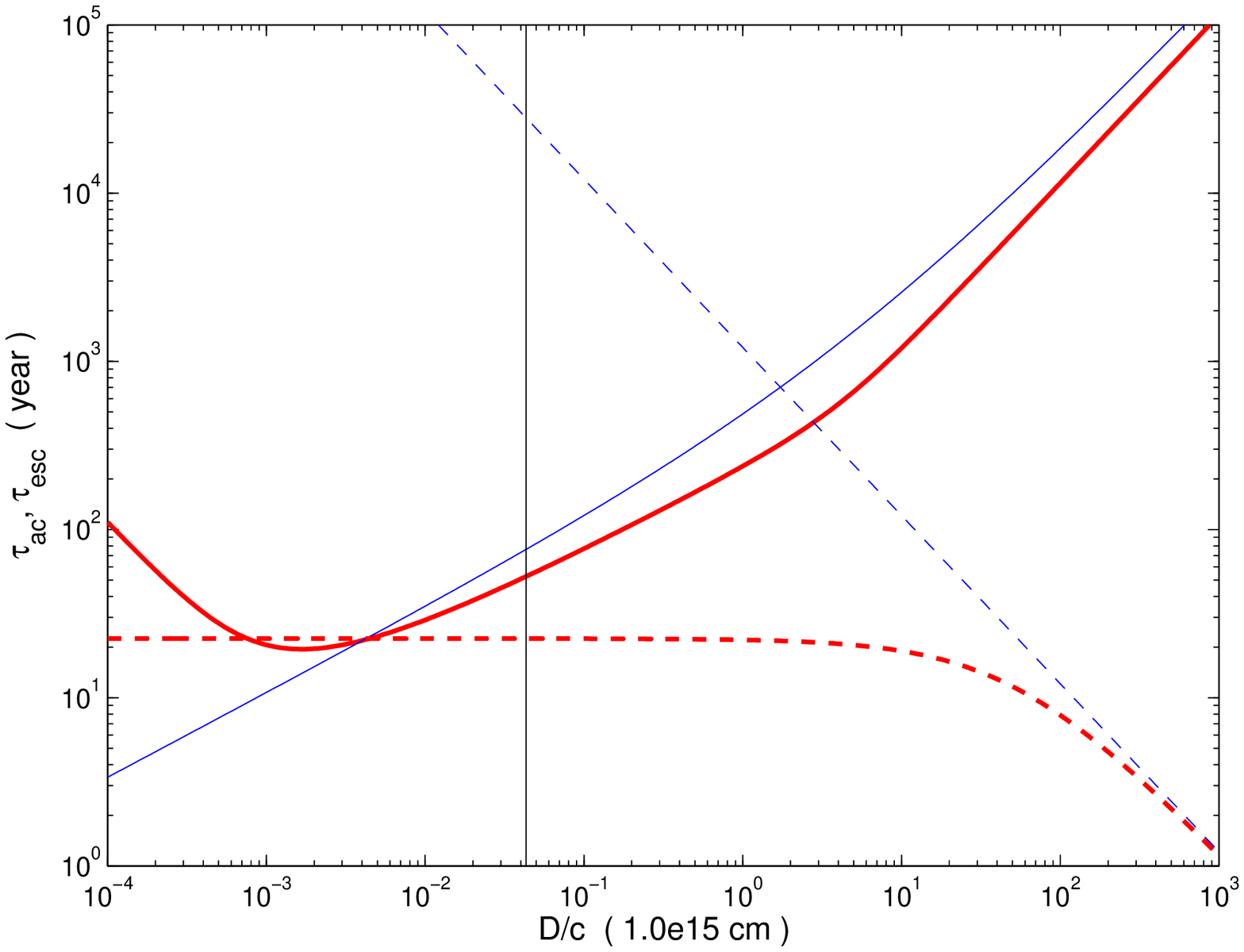}
\includegraphics[height=.27\textheight,width=0.48\textwidth]{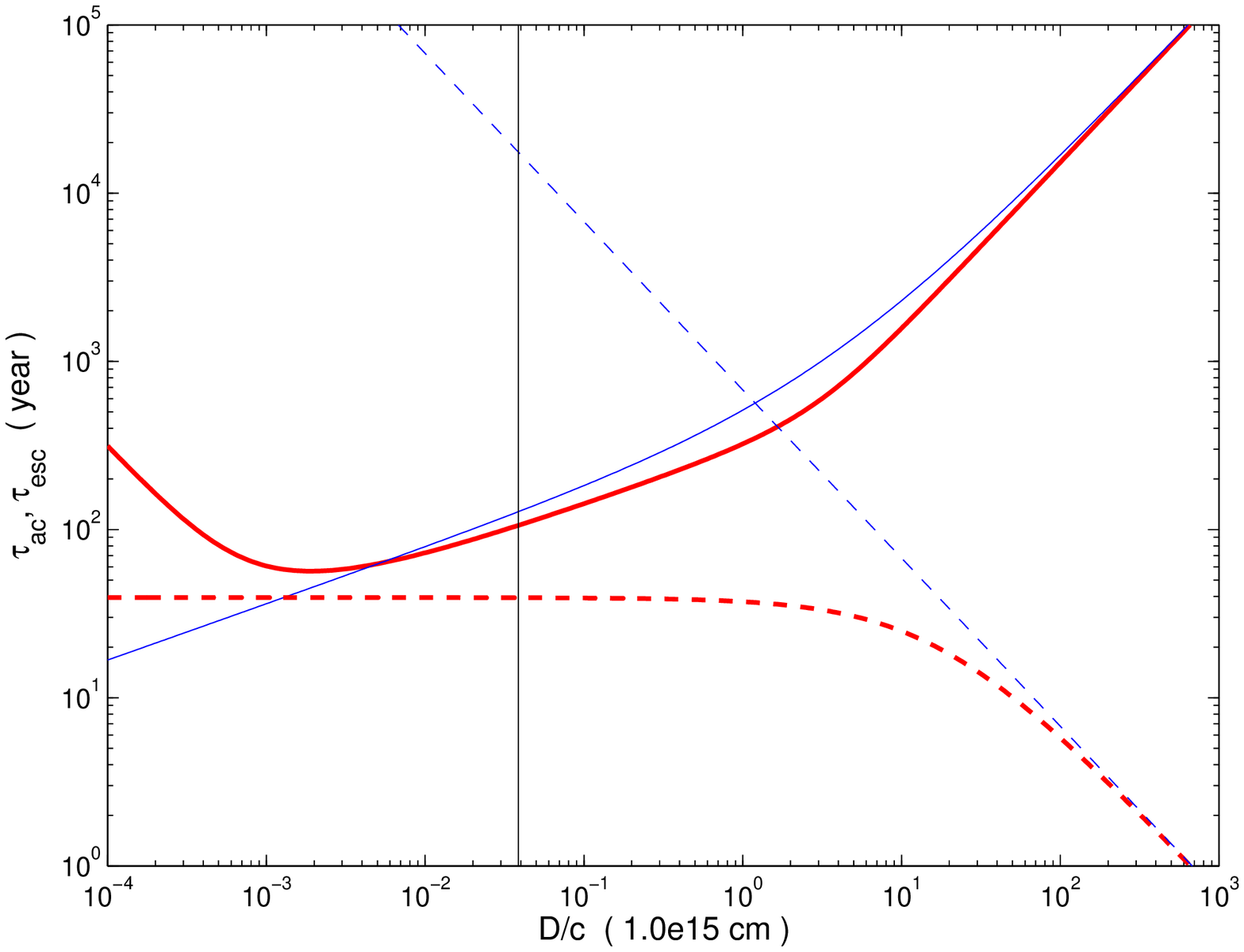}
\caption{The acceleration (solid) and escape (dashed) timescales as
a function of the diffusion coefficient $D$ for conditions at the
transonic point $x_1$ in the downstream, where $v_F=u$, for the
Kraichnan (left with $\chi=11$, $L= 10.7\times 10^{17}$cm and
$v_A=0.036U$) and Kolmogorov (right with $\chi=4.7$, $L = 8.0\times
10^{17}$cm and $v_A = 0.018U$) phenomenology in Figure \ref{vs}.
$U=4000$ km$/$s for both cases. The vertical line indicates the
diffusion coefficient $D = c l_d/3$. The thick lines are the exact
results given by Equations (\ref{tac}) and (\ref{tesc}). The thin
solid line indicates the approximate acceleration timescale given by
Equation (\ref{tac0}). The thin dashed line indicates
$(k_m^2D)^{-1}$. For a magnetic field of $B=14 \mu$G, the diffusion
coefficient for TeV electrons is about $10^{15}c$ cm. The
corresponding diffusion timescale over a length of $\sim3 \times
10^{16}$ cm, the half-width of the variable X-ray filements observed
with {\it Chandra} \citep{u07}, is about one years. \label{times}}
\end{figure}

The dependence of acceleration and escape timescales on $D$ at the
transonic point in the downstream for typical conditions of SNR RX
J1713.7-3946 are shown in Figure \ref{times}. The approximate
acceleration timescale given by Equation (\ref{tac0}) and the escape
timescale due to the diffusion coefficient $D$ alone, $(k_m^2D)^{-1}$,
are indicated with the thin lines.  At high values of $D$, the slightly
high discrepancy in the exact and approximate acceleration timescales
of the Kraichnan phenomenology is due to the fact that the overall wave
intensity is given by $2u^2$, instead of $3u^2/2$ as is for the
Kolmogorov phenomenology.

When $Dk_m\gg v_F$, corresponding to the fast diffusion limit,
$\tau_{ac}\simeq 3D/u^2$ and $\tau_{esc}\simeq (k_m^2D)^{-1}$. Then
$\tau_{ac}/\tau_{esc}\simeq 3D^2 k_m^2/u^2\gg 1$. The acceleration
is negligible. With these asymptotic expressions for these
timescales, the ratio of the acceleration to escape timescales
decreases with the decrease of $D$ and becomes close to unity near
$D\sim u/(3^{1/2}k_m)$. Since deviations from these asymptotic
expressions occur near $D\sim v_F/k_m\ge u/k_m$, efficient particle
acceleration is only possible in the regime where $D<v_F/k_m$. When
$Dk_d\ll v_F$, corresponding to the slow diffusion limit,
$\tau_{ac}\simeq 3(5-\delta)v_F^2/[(\delta-3) D k_d^{5-\delta}
k_m^{\delta-3}u^2]$.
\begin{equation}
\tau_{ac}=  \left[{(\delta-3)u^2 \over
          3D}\left({Dk_m\over v_F}\right)^{\delta-3}
          \int_{k_mD/v_F}^{k_dD/v_F}dx {x^{4-\delta}\over
         1+ x^2}\right]^{-1}
\simeq  \left[{(\delta-3)u^2 \over
          3D}\left({Dk_m\over v_F}\right)^{\delta-3}
          \int_{0}^{\infty}dx {x^{4-\delta}\over
         1+ x^2}\right]^{-1}
\sim u^{-2}D(Dk_m/v_F)^{3-\delta}\,,
\label{tac1}
\end{equation}
where $3<\delta<5$, which ensures the convergence of the
integration. These results are in agreement with Figure \ref{times}.

The acceleration and escape timescales are determined by $D$ and the
turbulence spectrum. Since we assume that $D=l_d c/3$ for electrons with
$r_g<l_d$, the electron acceleration and escape timescales are independent
of the energy for $r_g<l_d$. When the ratio of acceleration to escape
timescale is independent of the particle energy, a power-law particle
distribution is expected in the steady-state with the spectral index
given by \citep{p88}
\begin{eqnarray}
p & = & \left({9\over4} + {\tau_{ac}\over
\tau_{esc}}\right)^{1/2}-{1\over 2} \,. \label{index}
\end{eqnarray}
Since electrons with an energy less the proton rest energy can be
scattered by whistler waves at small scales, the diffusion
coefficient $D$ for these particles can be much less than $l_d c/3$
and the corresponding spectral index $p\simeq 1$  \citep{pl04, l06}.
The scatter will be dominated by the turbulent magnetic field before
electrons reach the proton rest mass energy, the spectral break may
appear below the proton rest mass energy. Without detailed treatment
of these processes, we will assume that the electron distribution
follows a power law with $p=1$ for $\gamma\le 10$ in the
following.\footnote{Assuming a spectral break at $\gamma=m_p/m_e$
will lead to a spectral bump near this break energy, in conflict
with radio observations. This suggests that, when the acceleration
time is long and $p$ is high, one needs better treatment of the
 electron acceleration at low energies.}
Above $\gamma=10$, the steady-state spectral index of the electron
distribution is given by Equation (\ref{index}) since the diffusion
coefficient $D=l_dc/3$ is assumed to be independent of the electron
energy.  For relativistic electrons with $r_g\ge l_d$, $D\ge r_g
c/3$. The increase of $D$ with the electron energy $E$ will lead to
softening of the electron distribution toward higher energy dictated
by the energy dependence of $\tau_{ac}/\tau_{esc}$.
\begin{figure}
\includegraphics[height=.27\textheight,width=0.48\textwidth]{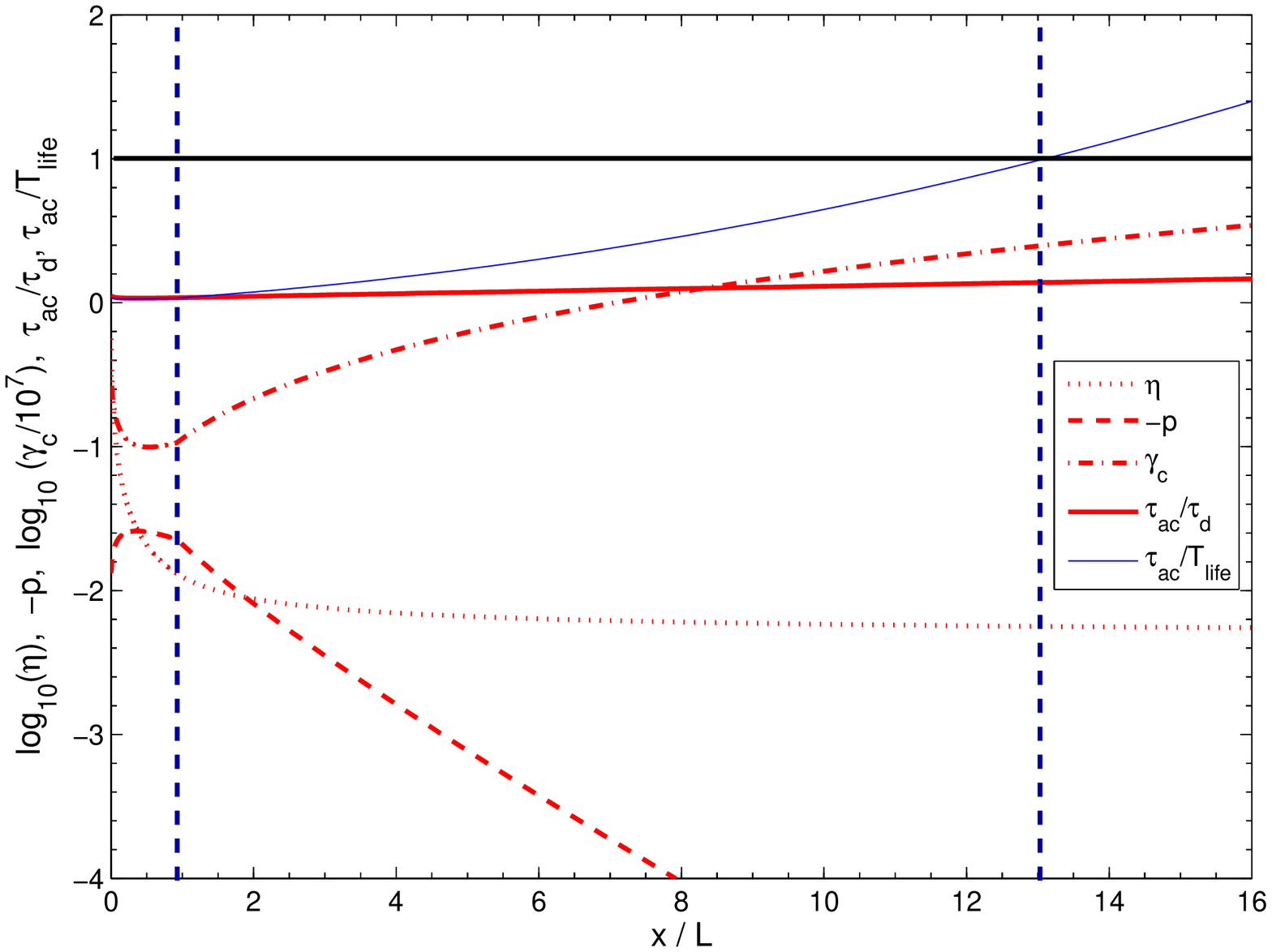}
\includegraphics[height=.27\textheight,width=0.48\textwidth]{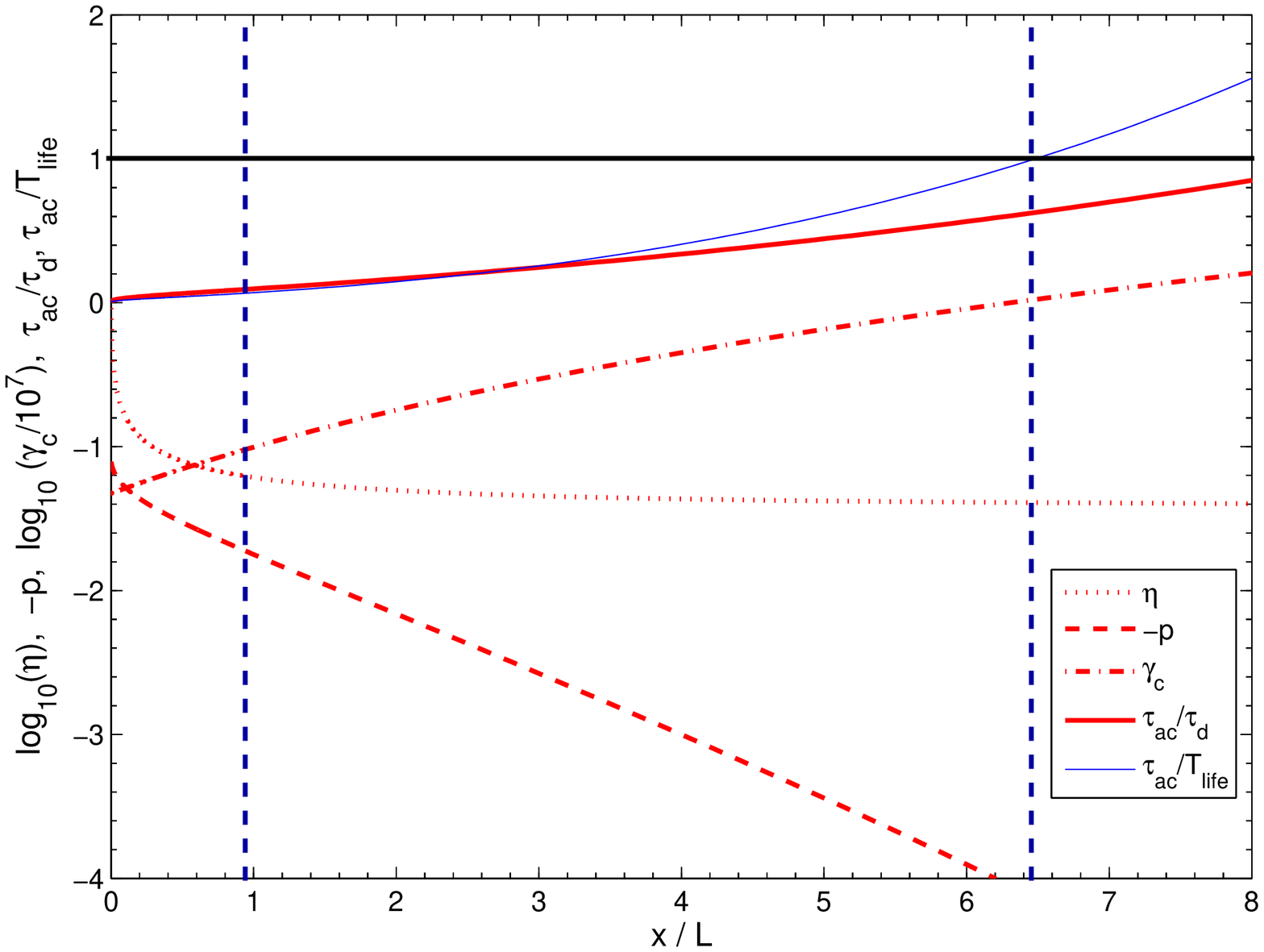}
\caption{Evolution of the acceleration efficiency $\eta$ (dotted),
cutoff Lorentz factor $\gamma_c$ (dotted-dashed), spectral index $p$
(dashed), and $\tau = \tau_{\rm ac}/T_{\rm life}$ (thin solid) in
the downstream for the Kraichnan phenomenology with $v_{A}=0.036U$
and $\chi=11$ (left) and Kolmogorov phenomenology with $v_A=0.018U$
and $\chi=4.7$ (right). $U=4000$ km/s. The particle acceleration is
significant for $\tau< 1$. We only consider acceleration between the
two vertical dashed lines indicating $x_1$ and $x_2$. For
$\gamma_c$, we have assumed that $B=14\mu$G, $L=10.7\times10^{17}$
cm (left) and $B=14\mu$G, $L=8.0\times 10^{17}$ cm (right).
\label{paras}}
\end{figure}
\begin{figure}
\includegraphics[height=.27\textheight,width=0.48\textwidth]{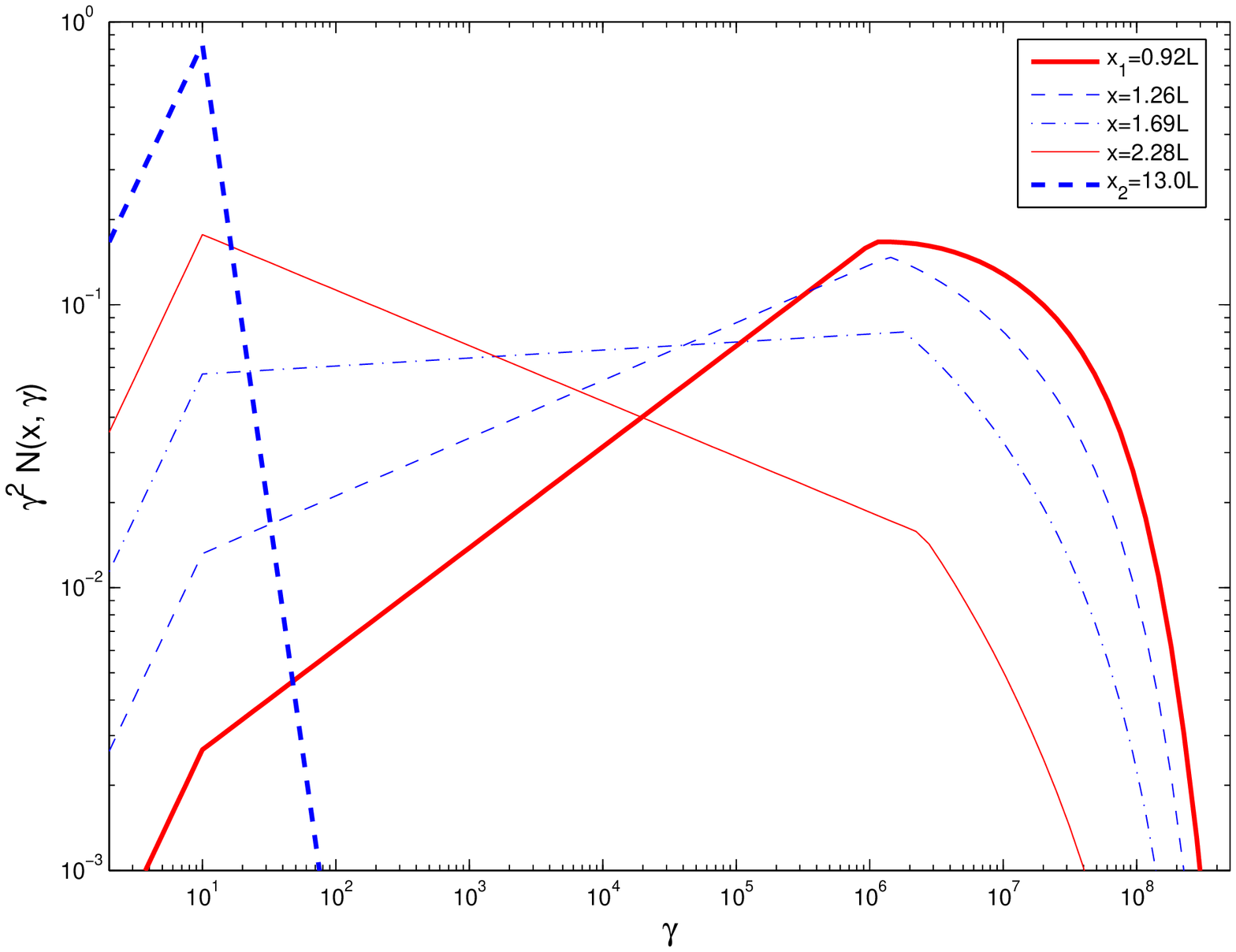}
\includegraphics[height=.27\textheight,width=0.48\textwidth]{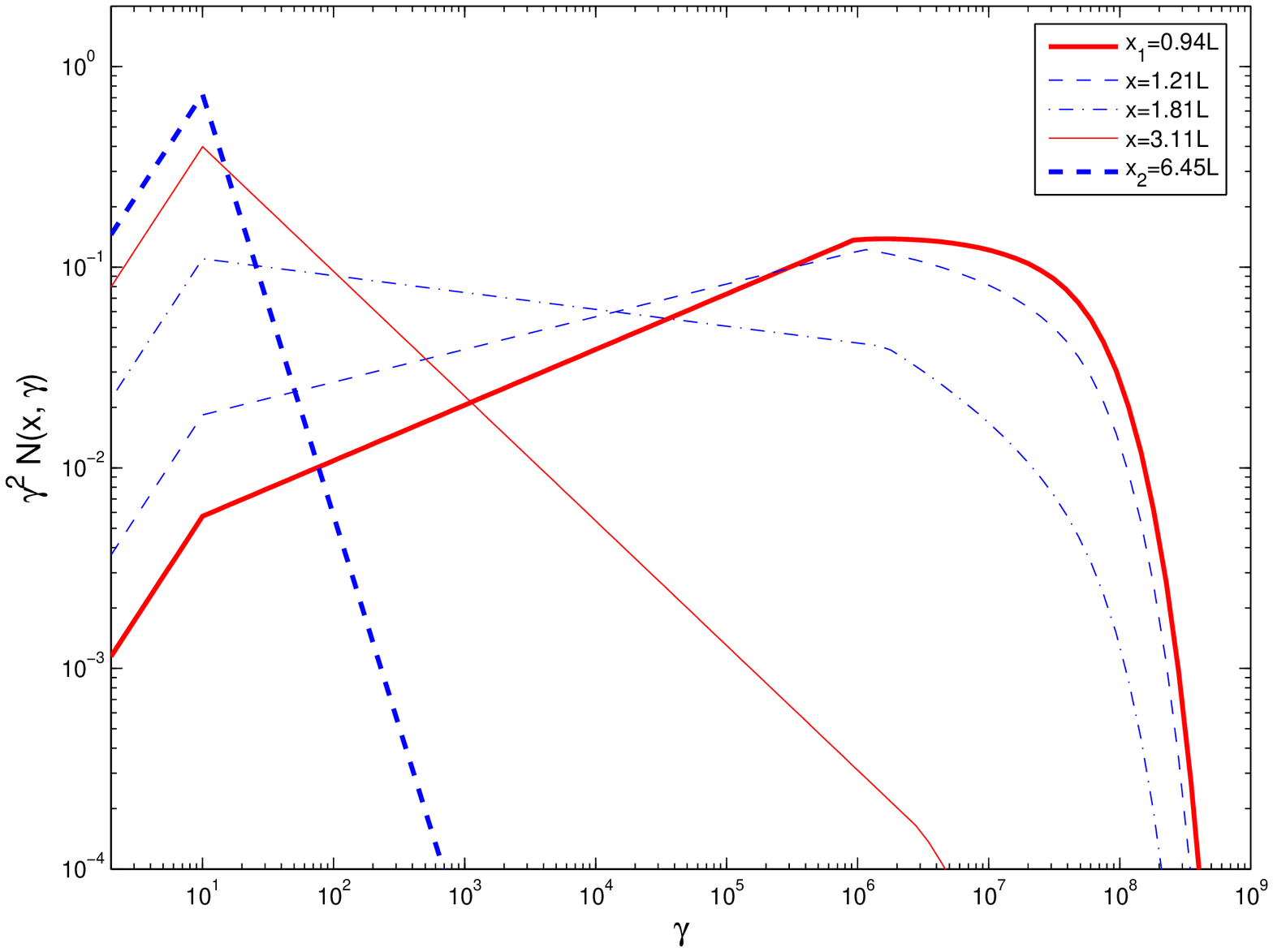}
\caption{ Normalized steady-state nonthermal electron distribution
$f(x)$ produced at several locations in the downstream for the Kraichnan
(left) and Kolmogorov (right) phenomenology in Figure \ref{paras}.
\label{f}}
\end{figure}
The exact electron distribution can be obtained numerically. For the
sake of simplicity, we assume that the electron distribution has a
high-energy cutoff at Lorentz factor
\begin{equation}
\gamma_c={qB\over m_ec^2k_d} = {qBL v_A^3\over m_e c^2 u^3}
\left\{
\begin{array}{ll}
v_A/\min(v_F, u)
\ \ \ \ \ &{\rm for\ IK}\,, \\
1\ \ \ \ \ &{\rm for\ Kol}\,.\\
\end{array}
\right. \label{cutoff}
\end{equation}
where $q$ is the elementary charge units. Then for a steady-state
treatment, the distribution of electrons escaping from the
acceleration site may be approximated reasonably well with
\citep{bld06, pp95}
\begin{equation}
f(x, \gamma)\propto \gamma^{-p(x)} \exp
\{-[\tau_{ac}(D)/\tau_{esc}(D)]^{1/2}\}/\tau_{esc}(D)\,, \label{f1}
\end{equation}
where
\begin{equation}
D(\gamma, x)=
\left\{
\begin{array}{ll}
c l_d(x)/3 \ \ \ \ \ &{\rm for\ \gamma\le \gamma_c(}x)\,, \\
c r_g(\gamma)/3 = \gamma m_e c^3/(3 qB)\ \ \ \ \ &{\rm for\ \gamma>
 \gamma_c(}x)\,.\\
\end{array}
\right.
\end{equation}
We note that the shape of the distribution function near $\gamma_c$
can also be affected by the energy loss processes \citep{s08, v09, b10}.
In our model, it is mostly determined by the balance between the
acceleration and escape processes \citep{pp95, bld06, z07}. The shape of
this high-energy cutoff has significantly effect on the fitting
parameter, especially the value of $\gamma_c$. We are carrying out
detailed numerical investigation of the particle distribution and the
results will be reported in a separate publication.

Although the parallel propagating waves in the dissipation range
may not contribute to the particle scatter significantly, energies carried
by these waves are only accessible to relativistic electrons. At
very small scales, these waves resonate with thermal background
electrons giving rise to a preferential acceleration of electrons.
One therefore may assume that the ratio of the
dissipated energy carried by non-thermal electrons to that of the
thermal ions is proportional to the ratio of the energy density
of parallel propagating fast-mode waves to that of the magnetic field:
\begin{equation}
\eta = {\theta_c^2(k_d)\over 2} = \left\{
\begin{array}{ll}
{e^{5/6}v_A^2/ (2\pi)^{1/2} v_S
v_F}\ \ \ \ \ &{\rm for\ IK}\,, \\
{e^{5/6}v_A/ (2\pi)^{1/2} v_S}\ \ \ \ \ &{\rm for\ Kol}\,.\\
\end{array}
\right. \label{efficiency}
\end{equation}
where $e\simeq 2.72$ is the base of the natural logarithm. A more
quantitatively treatment of this issue may address the electron
injection processes self-consistently \citep{e79b}.

To have efficient acceleration of relativistic electrons, the
turbulence decay time
\begin{equation}
\tau_d =3L/C_1u \left\{
\begin{array}{ll}
\max(u, v_F)/u\ \ \ \ \ &{\rm for\ IK}\,, \\
1\ \ \ \ \ &{\rm for\ Kol}\,.\\
\end{array}
\right.
\end{equation}
and the remnant lifetime $T_{\rm life} $ should be longer than the
acceleration time. As we will see below, the turbulence decay time
is always longer than $T_{\rm life}$ in the subsonic phase for SNR
RX J1713.7-3946. There are at most two locations $x_0$ and $x_2$
with $x_0<x_2$ in the downstream, where $\tau=\tau_{\rm ac}/T_{\rm
life}=1$. Figure \ref{paras} shows the evolution of $\eta$,
$\gamma_c$, $p$, $\tau_{\rm ac}/\tau_{\rm d}$ and $\tau=\tau_{\rm
ac}/T_{\rm life}$ in the downstream for $U=4000$ km/s. The profiles
of $v_F/U$ and $u/U$ only depend on $v_A/U$. So is the profile of
$\eta$. The profiles of $\tau$ and $p$ also depend on the absolute
value of $U$. To obtain $\gamma_c$, one needs to know $L$ and $B$ as
well. Most SA occurs near the sonic point $x_1$, where $v_F=u$. In
the late subsonic phase, $u\ll v_F$, the SA is insignificant since
most of the free energy of the system has been converted into heat.
The characteristic length of the magnetic field is also long far
downstream due to the weak turbulence, which implies long electron
scatter and acceleration timescales.

\begin{figure}
\includegraphics[height=.27\textheight,width=0.48\textwidth]{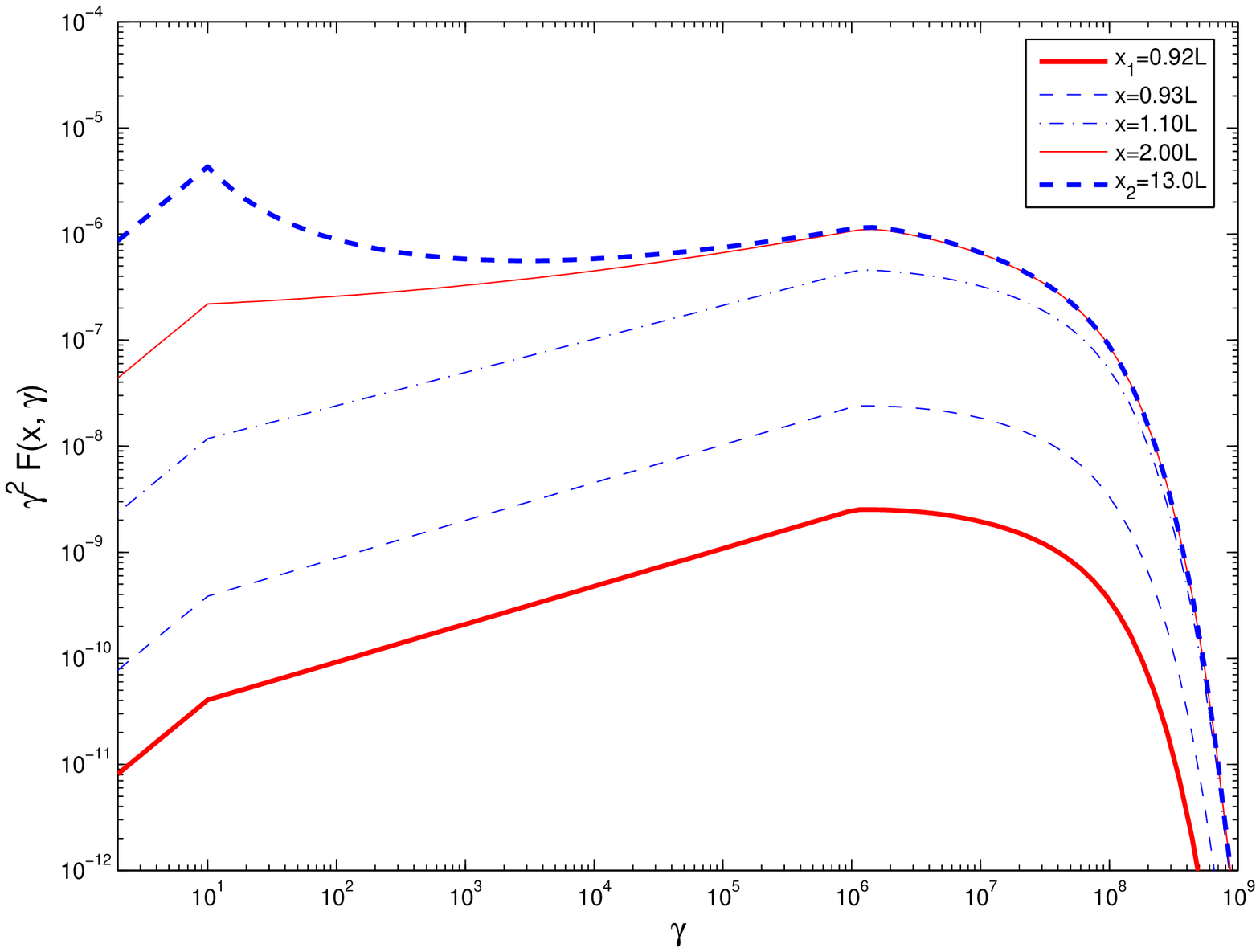}
\includegraphics[height=.27\textheight,width=0.48\textwidth]{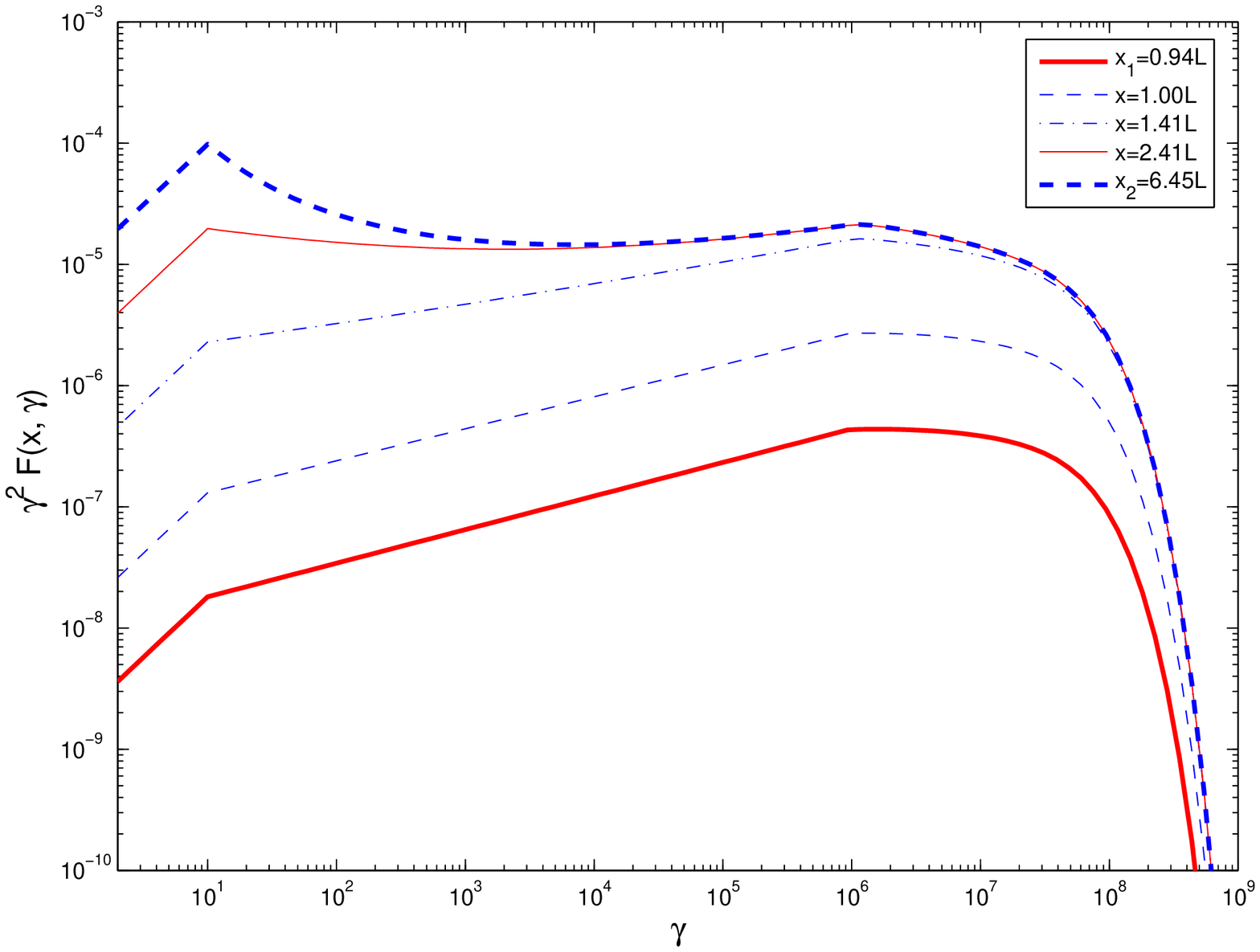}
\caption{Distributions of nonthermal electrons $F(x)$ at several
locations in the downstream for the Kraichnan (left) and Kolmogorov
(right) phenomenology in Figure \ref{paras}. \label{Fy}}
\end{figure}

Then the distribution of non-thermal electrons in the downstream
\begin{equation}\label{dis}
F(x, \gamma) = \int_{x_1}^x f(x^\prime, \gamma) \eta(x^\prime)
(4Q/m_ec^2U) {\rm d} x^\prime \label{F0}
\end{equation}
where $\int_{1}^\infty \gamma f(x^\prime, \gamma){\rm d}\gamma
= 1$, and $\int_{1}^\infty \gamma m_e c^2F(x, \gamma){\rm
d}\gamma$ gives the energy density of non-thermal particles at $x$.
Figures \ref{f} and \ref{Fy} show the normalized electron
distribution $f$ and $F$ for parameters in Figure \ref{paras} at
several locations in the downstream, respectively. We note that the
electron distribution has a rather gradual high-energy cutoff due to the
weak dependence of the acceleration and escape timescales on the spatial
diffusion coefficient $D$, which is proportional to the electron energy
above the cutoff energy $\gamma_c m_ec^2$. This gradual cutoff results
in a broad emission
component due to inverse Comptonization of the low-energy background
photons by high-energy electrons, which can fit the observed broad TeV
emission spectrum from a few SNRs.

\section{Application to SNR RX J1713.7-3946 and Time-Dependent Models}
\label{application}

\begin{figure}
\includegraphics[height=.27\textheight,width=0.48\textwidth]{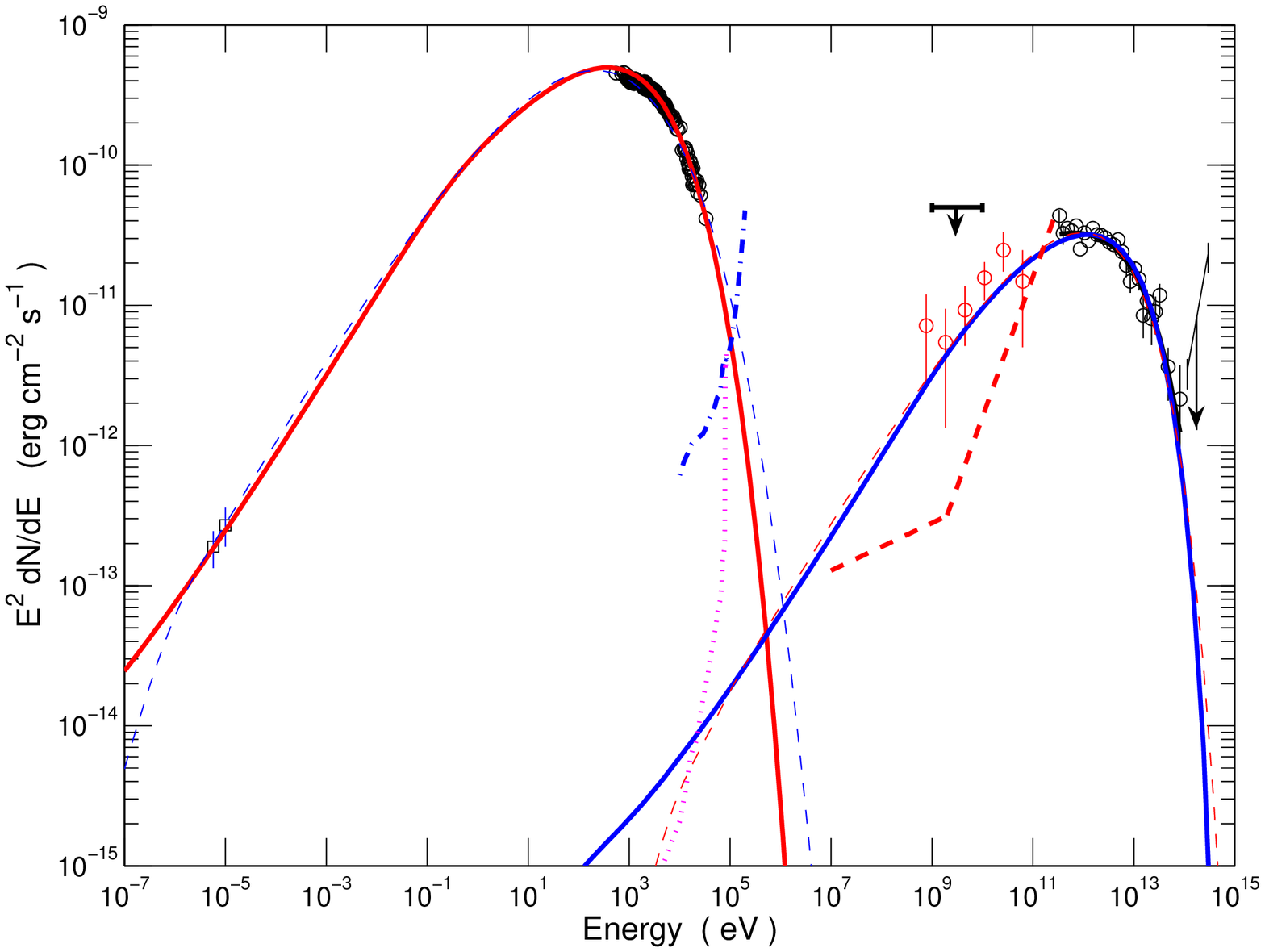}
\includegraphics[height=.27\textheight,width=0.48\textwidth]{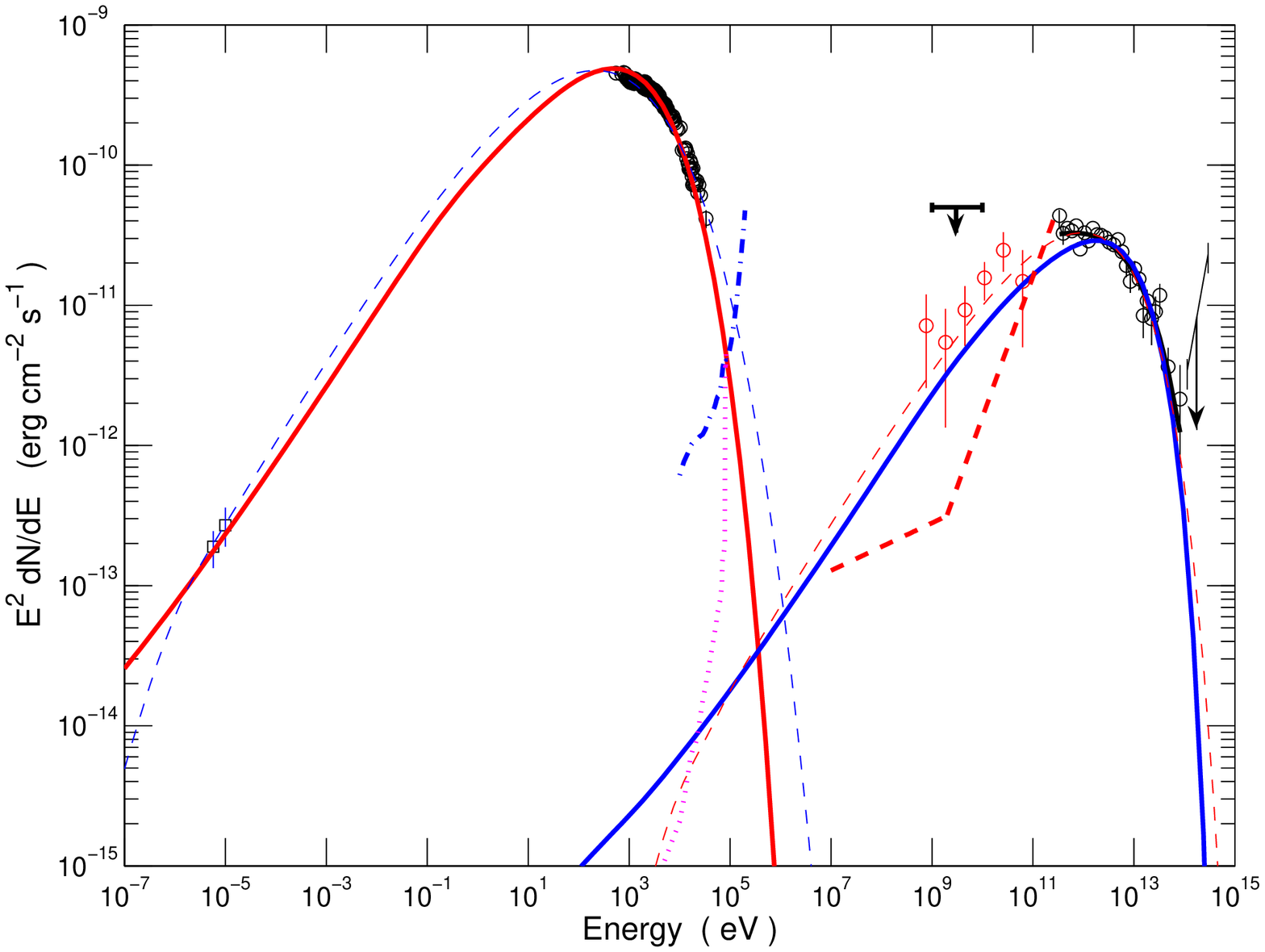}
\caption{ Best fits to the observed spectrum of  the SNR RX J1713.7-3946.
The X-ray data points are obtained from \citet{t08}. The other data
points and the sensitivity limits of different high-energy telescopes
are the same as those in \citet{l08}. The left and right panels are
for the Kraichnan and Kolmogorov phenomenology in Figure \ref{paras},
respectively. The dashed line is for a simple power-law model with
a gradual high-energy cutoff $F(\gamma)\propto \gamma^{1.85}
\exp(\gamma/\gamma_c)^{1/2}$, where $\gamma_cm_ec^2 = 3.68$ TeV. The
solid lines are for the fiducial models. The low and high energy
spectral peaks are produced through the synchrotron and inverse Compon
scatter of the background photons \citep{p06}, respectively. The
preliminary data from Fermi were not considered in the fit and are
included here to show the agreement between model predictions and this
observation \citep{funk09}. The Fermi observation seems to favor the
Kraichnan model. \label{spec}}
\end{figure}

Here we use the SNR RX J1713.7-3946 as an example to demonstrate how
the SA by fast-mode waves accounts for the observed broadband
spectrum. The electron distribution produced by the SA is given by
equation (\ref{F0}), where the integration over $x^\prime$ should
stop at $x_2$. Given the evolution history of the SNR, the
nonthermal electron distribution will vary along the radial
direction. A detailed modeling of the explosion is necessary to take
this effect into account properly \citep{cs84, z10}. Here we treat
the volume of the emission region $V_e$ as a free parameter to
control the normalization of the emission spectrum, which is
appropriate for SNRs where the nonthermal particles appear to be
concentrated near the SF. By adjusting $U$, $B$, $v_A$, $\chi$, $L$,
and $V_e$, one can use the corresponding electron distribution
$F(x_2)$ to fit the observed spectrum of the SNR RX J1713.7-3946.
Figure \ref{spec} shows the best fit with the corresponding
parameters listed in Table \ref{t1}, where $V$ is the enclosed
volume of the SNR SF. Comparing to the thin-dashed line, which is
derived by assuming an electron distribution $\propto
\gamma^{-p}\exp-(\gamma/\gamma_c)^{1/2}$, the high-energy cutoff of
the Kraichnan model is more gradual than the Kolmogorov model, which
makes the former spectrum broader. The emission volume is $0.42$
($0.016$) times the volume of the SNR for the Kraichnan (Kolmogorov)
phenomenology, which is compatible with observations. The values of
$\chi$ are greater than 1, which suggests that the particle spatial
diffusion coefficient might be enhanced significantly by
incompressible turbulence motions. The total energies of nonthermal
electrons are also comparable to the magnetic field energy for a
uniform magnetic field within the remnant, suggesting near energy
equipartition between the nonthermal electrons and the magnetic
field.

\begin{table*}
\caption{Model Parameters} \centering
\begin{tabular}{lcccccccc}
\hline
model & U & B & $v_A/U$ & $\chi$ & L & $V_e/V$ & $E_{e}$ & $\rho$\\
      & (km s$^{-1}$) & ($\mu$G) & & & $10^{17}$(cm) &  & $10^{47}$(erg) & $10^{-26}$(g
      cm$^{-3}$)\\
\hline
IK steady          & 4000 & 14.0 & 0.036 & 11  & 10.7 & 0.42  & 7.69 & 7.52 \\
KOL steady         & 4000 & 14.0 & 0.018 & 4.7   & 8.0  & 0.016 & 6.96 & 30.1 \\
\hline
\end{tabular}
\label{t1}
\end{table*}

There are six parameters in the model: $B$, $U$, $v_{A}$, $\chi$,
$L$, and $V_e$. The observed radio to X-ray spectral index, X-ray to
TeV flux ratio, location of the X-ray cutoff, and bolometric
luminosity of the source give four constraints, which leads to two
more degree of freedom. Our model fit to the spectrum therefore is
not unique. $\chi$ is determined by the coupling between
incompressible and compressible turbulence motions. $B$ is well
constrained by the ratio of the X-ray to TeV flux. To reproduce the
observed spectral shape, the profiles of $p$, $\gamma_c$, and $\eta$
should not change significantly, which implies that $v_A^8c^2\propto
u^{10}$ and $L\propto u^4/v_A^4$ at the transonic point for the
Kraichnan phenomenology. For the Kolmogorov model, $v_A^6c^2\propto
u^{8}$ and $L\propto u^3/v_A^3$. For $v_A\ll U$, $u$ is proportional
to $U$, we find that nearly identical emission spectra can be
obtained by adjusting $U$ and $v_A$ \citep{l08b}. Since the
turbulence decay time is longer than the supernova lifetime, the
steady-state treatment is also justified. A time-dependent treatment
gives identical parameters \citep{bld06}.

\section{Density and Turbulence Generation}
\label{disc}

The primary discrepancy between these models and the observations
are the relatively high densities of the downstream plasma. From
X-ray observations, \citet{c04} inferred an upper limit for the
electron density of 0.02 cm$^{-3}$. The corresponding mass density
is about $3.3\times 10^{-26}$ g cm$^{-3}$, which is comparable to
the densities inferred with the Kraichnan phenomenology but lower
than those inferred with the Kolmogorov phenomenology. On the other
hand, the electron temperature could be much lower than the ion
temperature in the shock downstream \citep{z10}. \citet{m09} and
\citet{f09} have argued that the density in the downstream can be as
high as $0.5$ cm$^{-3}$, corresponding to a mass density of
$8.4\times 10^{-25}$g cm$^{-3}$.

Given the age of $T_{\rm life}=1600$ years and a radius of $R=10$ pc
at a distance of $D\simeq 1$ kpc, the corresponding average speed of
the shock front is $6100$ km s$^{-1}$.  With the self-similar
solution of \citet{c82}, we infer a shock speed of
$[(n-3)/(n-s)]6100$ km s$^{-1}$, where $n>5$ and $0\le s<3$ are the
power-law exponents of the density profile for the ejecta and the
ambient medium, respectively. Observations give an upper limit of
$4500$ km s$^{-1}$for the shock speed \citep{u07}. From the
self-similar solution, the shock speed must be higher than $2400$ km
s$^{-1}$ (for $s=0$ and $n=5$). If ions are preferentially heated by
the shock, the corresponding ion temperature $T_i$ will be higher
than $3m_pU^2/16k_{\rm B}>1.3\times 10^8$ K. The electron
temperature should be higher than that given through Coulomb
collisional energy exchange with ions \citep{h00}:
\begin{equation}
T_e>2.1\times 10^7 (T_{\rm life}/1600{\rm yr})^{2/5}(n_e/{\rm
cm}^{-3})^{2/5}(T_i/1.3 \times 10^8{\rm K})^{2/5}{\rm K}\,,
\end{equation}
where $n_e$ is the electron number density. The corresponding
bremsstrahlung luminosity is ${\cal L} > 5.2\times
10^{34}(n_e/0.5{\rm cm}^{-3})^{11/5}$ erg s$^{-1}$, which is
comparable to the luminosity of the observed nonthermal X-ray
emission.\footnote{Here we have assumed an emission volume one
quarter of the volume enclosed by the remnant shock front.} We
therefore expect strong thermal emission with such a high density.
\citet{m09} obtained a very low thermal bremsstrahlung luminosity by
arbitrarily adopting an electron temperature 100 times lower than
the ion temperature. As shown above, considering the electron ion
Coulomb collisional energy exchange, the electron temperature will
not be that low and significant thermal X-ray is expected with a
density of $0.5$ cm$^{-3}$, except that cooling of the shock front
by cosmic ray ions dominates \citep{z10}. The highest electron
density given by our models is about $0.2$ cm$^{-3}$. The
corresponding thermal X-ray luminosity will be reduced by nearly one
order of magnitude and should be in agreement with observations.
Detailed modeling of the supernova explosion and the thermal
emission is needed to see the validity of these models.

The model inferred density may also be reduced by considering the
acceleration of electrons by large-scale structures in the
downstream and the acceleration in the supersonic phase, where the
first-order Fermi acceleration is also possible. In this paper, we
consider the electron acceleration by the fully developed turbulence
in the subsonic phase. It assumes that once the electrons diffuse
over a scale of the turbulence generation length $L$, the
acceleration stops. As shown above, the turbulence evolves in the
downstream. In a more self-consistent treatment, one may use the
turbulence properties to derive nonthermal electrons injected into
the downstream flow by small scale plasma waves and consider the
further acceleration of these electrons as they diffuse spatially in
the downstream. The scatter mean-free-path of these particles are
determined by the properties of turbulence. The electron
acceleration stops only after they diffuse into upstream or far
downstream, where the turbulence becomes insignificant. If these
effects lead to a harder overall electron distribution, the
Alfv\'{e}n speed needs to be increased to fit the observations,
leading to a lower density.

From these models studied here, we see that, to have efficient SA,
both high-speed waves ($v_F\simeq u$) and short scatter
mean-free-path are required. Quantitatively, one needs
$c^2l^2/L^2u^2$ to be on the order of unity so that the acceleration
and escape time scales of relativistic particles are comparable. $u$
is constrained by the shock speed. A short scatter mean-free-path is
achieved by the reduction of the characteristic length of the
magnetic field, which also determines the maximum energy of the
accelerated particles. In these models, turbulence motions are
invoked to reduce the characteristic length of the magnetic field.
It is obvious, such a mechanism is only possible for strong
turbulence, where the turbulence speed is higher than the Alfv\'{e}n
speed. The turbulence speed is determined by the shock speed $u\le
(3/16)^{1/2}U$, which is less than 1949 km s$^{-1}$ for SNR RX
J1713.7-394. Therefore $v_A<1949$ km s$^{-1}$ and we obtain a low
limit for the mass density from the inferred magnetic field of 14
$\mu$G: $\rho = B^2/4\pi v_A^2> 4.1\times 10^{-28}$g
cm$^{-3}(B/14{\mu \rm G})^2 (U/4500{\rm km\ s}^{-1})^{-2}$, which
corresponds to an electron density of $\sim 0.0002$
cm$^{-3}(B/14{\mu \rm G})^2 (U/4500{\rm km\ s}^{-1})^{-2} $.

Therefore further reduction of the density can be achieved by
considering the generation of the turbulence and its effect on the
turbulence spectrum, i.e., the dependence of the eddy velocity on
the spatial scale. Indeed, in the models considered above, we assume
that the turbulence is generated in a very narrow spatial range
instantaneously at the SF and an inertial range develops. Since the
large-scale eddy speed is comparable to the bulk velocity of the
downstream flow moving away from the shock front, the region with
$x<0.5$ should be considered as the turbulence generation phase.
This is an intrinsic limitation of the above treatments, which focus on
the averaged properties of the downstream flow without
addressing the turbulence generation process. Significant particle
acceleration occurs within $x<0.5$, i.e., the turbulence generation
phase. It is also possible that the turbulence is generated over a
broad spatial range and/or the turbulence is not isotropic at large
scales. One then expects a turbulence spectrum shallower than the
initial range spectrum. For example, for a turbulence spectrum of
\begin{equation}
W= [3u^2/8\pi(g-3)] (2\pi/L)^{g-3} k^{-g-2}
\end{equation}
with $1<g<1.5$, \footnote{For $g<1$, the eddy speed increases with
the decrease of the spatial scale and the turbulence energy is not
dominated by the large scale eddies. One then needs to introduce a
spectral break at certain small scale. Such a complex scenario is
not well justified from both observational and theoretical point of
views.} where the normalization is chosen so that the turbulence
energy density is given by $(3/2)\rho u^2$, the eddy speed can be
redefined as
\begin{equation}
v_{edd}\equiv \{[8\pi(g-1)/3] k^3 W\}^{1/2} = u
(kL/2\pi)^{(1-g)/2}\,.
\end{equation}
The eddy speed is comparable to the Alfv\'{e}n speed at the
characteristic length of the magnet field $l$, we then have
$12c^2l^2/v_f^2L^2 = 12 c^2v_A^{4/(g-1)}/v_f^2u^{4/(g-1)}\sim 1$.
Therefore $v_A/u \sim (v_f/12c)^{(g-1)/2}$. Since $v_f\sim u\sim U$,
we have $v_A/U\sim (U/c)^{(g-1)/2}$. The Alfv\'{e}n speed can be
comparable to the turbulence speed for a shallow turbulence spectrum
with $g$ approaching 1. Thus stochastic electron acceleration can
account for observations of SNR RX J1713.7-394 as far as the mass
density of the shocked plasma is greater than $4.1\times 10^{-28}$g
cm$^{-3}$. Detailed studies of the turbulence generation and the
associated particle acceleration are warranted \citep{lb00, h04,
gj07, n09}.

\section{Summary and Conclusions}
\label{con}

In the paper, we study the SA of electrons by a decaying turbulence
as produced and/or enhanced by strong non-relativistic shocks and
carried away from the SF with the downstream flow. It is shown that,
to have significant particle acceleration, the turbulence must cover
a large spatial scale so that the particle acceleration time may be
shorter than the turbulence decay time. To account for observations
of a few STTSNRs with the leptonic scenario for the TeV emission,
fast-mode waves need to be excited in the subsonic phase. Given
the turbulent nature of the downstream flow, fast-mode waves may
prevail in the downstream. We show that the SA by large-scale acoustic
(fast-mode) waves can account for the observations.

There are four basic model parameters, namely, the magnetic field,
mass density, shock speed, and the turbulence generation scale.
Observations of a few SNRs and radio galaxies have shown that the
particle acceleration may change dramatically along the SF
\citep{r04, r09, c09}. This variation has been attributed to a
large-scale magnetic field in the DSA model as the acceleration
efficiency varies with the angle between the magnetic field and the
shock normal. With our model, this variation is likely caused by a
quite different mechanism. Detailed comparative studies should be
able to distinguish these models.

The particle acceleration is very sensitive to the magnetic field.
To produce nonthermal particle distribution in compatible with
observations, $c^2v_A^8/u^{10}$ and $c^2v_A^6/u^8$, where $u$ is the
large-scale eddy speed near the shock front, should be on the order
of 1 for the Kraichnan and Kolmogorov phenomenology, respectively.
The high-energy cutoff of the particle distribution is determined by
the magnetic field and the turbulence generation scale. Weaker
fields will lead to lower cutoff energies and harder spectra. The
particle acceleration may be turned off completely for strong fields
due to the increase of the characteristic length of the magnetic
field, and therefore the particle scatter mean-free-path.
If the magnetic field is predominantly generated by the streamline
of nonthermal particles upstream, the models then imply $v_A\sim
U^{5/4}$ and $v_A\sim U^{4/3}$ for this dynamo process and for the
Kraichnan and Kolmogorov phenomenology in the downstream turbulence,
respectively.

Assuming that the turbulence is isotropic and generated in a narrow
spatial scale, the model inferred densities of the downstream flow
may be so high that thermal X-ray emission becomes
observable, in conflict with observations. Although the thermal
X-ray emission may be suppressed by a lower electron temperature due
to dominance of the cooling by cosmic ray ions, we find that a low
density is also possible if the turbulence is not isotropic or
generated over a broad spatial scale so that the eddy speed has very
weak dependence on the spatial scales. Detailed modeling of the
progenitor and the evolution history of the remnant may help to
constrain the density \citep{cs84}.

The application of the models to the SNR RX J1713.7-3946 also
suggests energy equipartition between the magnetic field and the
acceleration electrons. If the rest of the shock energy is
dissipated as heat in the downstream, then the overall electron
acceleration efficiency will be $\sim v_A^2/u^2$, which is inversely
proportional to the plasma $\beta$ of the downstream flow. For the
Kraichnan phenomenology, $v_A^2/u^2 \sim (u/c)^{1/4}$ and for the
Kolmogorov phenomenology $v_A^2/u^2\sim  (u/c)^{1/3}$, acceleration
is more efficient for stronger shocks, which also produce hotter
downstream plasma. However, the dependence of the acceleration
efficiency on the shock speed is rather weak. These may have
significant implications on the origin of cosmic rays and their
connection to the properties of the interstellar medium \citep{r08}.

With the fast-mode wave turbulence model studied in this paper,
relativistic electrons may also be accelerated through the first-order
Fermi mechanism as in the DSA model, especially in the supersonic
phase. The high-energy cutoff can
still result from decoupling of higher energy particles with the
background magnetic field as their gyro-radius exceeds the
characteristic length of the magnetic field. This is quite different
from the DSA models, where the high-energy cutoff is related to a
finite lifetime of the shock, shock curvature, or efficient energy loss
processes \citep{z10}.


\section*{Acknowledgements}
We thank the referee for critical reviews of the paper, which made
us examine the acceleration timescale and the shape of the
high-energy cutoff of the electron distributions carefully. SL
acknowledges support from the EU's SOLAIRE Research and Training
Network at the University of Glasgow (MTRN-CT-2006-035484) and
thanks Lyndsay Fletcher for constant support and John Kirk and
Christian Fendt for invitation to the workshop entitled ``The
high-energy astrophysics of outflows from compact object'', where
some ideas in this paper were conceived. This work is supported in
part under the auspices of the US Department of Energy by its
contract W-7405-ENG-36 to Los Alamos National Laboratory and by the
National Science Foundation of China (grants 10963004 and 10778702),
Yunnan Provincial Science Foundation of China (grant 2008CD061) and
SRFDP of China (grant 20095301120006).





\bsp

\label{lastpage}

\end{document}